\pdfoutput=1
\documentclass[11pt]{article}
\usepackage[utf8]{inputenc}
\usepackage{amsmath,amssymb,amsfonts}
\usepackage{mathtools,bm}
\usepackage{booktabs,multirow,array,longtable}
\usepackage{rotating,pdflscape}
\usepackage{xspace}
\usepackage{xcolor}
\usepackage{microtype}
\usepackage{tikz}
\usetikzlibrary{arrows.meta}
\usepackage[letterpaper,textwidth=6in,textheight=8.5in]{geometry}
\usepackage{graphicx}
\usepackage[round,semicolon,authoryear]{natbib}
\usepackage[hidelinks]{hyperref}
\usepackage{authblk}
\newtheorem{theorem}{Theorem}

\newtheorem{proposition}[theorem]{Proposition}

\newtheorem{example}{Example}
\newenvironment{keywords}{\begin{quotation}\small\noindent\textbf{Keywords:} }{\end{quotation}}
\usepackage[ruled]{algorithm2e}
\mathtoolsset{showonlyrefs=true}
\hypersetup{pdftitle={DAG-CLIP: A DAG Learning Framework in the Presence of Latent Variables},
  pdfauthor={Siliang Zhang, Yunxiao Chen, and Irini Moustaki}}

\newtheorem{assumption}{Assumption}

\newcommand{\argmin}{\operatornamewithlimits{arg\,min}}
\newcommand{\argmax}{\operatornamewithlimits{arg\,max}}

\newcommand{\diag}{\operatorname{diag}}
\newcommand{\tr}{\operatorname{tr}}

\title{DAG-CLIP: A DAG Learning Framework\\
in the Presence of Latent Variables}

\author[1]{Siliang Zhang}
\author[2]{Yunxiao Chen}
\author[2]{Irini Moustaki}
\affil[1]{School of Statistics, East China Normal University}
\affil[2]{Department of Statistics, London School of Economics and Political Science}
\date{}

\renewcommand{\baselinestretch}{1.2} 

\begin{document}
\maketitle

\begin{abstract}
  Learning directed acyclic graphs (DAGs) to uncover causal mechanisms has attracted substantial attention in machine learning. While most existing methods focus exclusively on observed variables, many variables of substantive interest are latent constructs defined by statistical measurement models. Such constructs are particularly common in the social and behavioral sciences. In this paper, we propose a general statistical framework for DAG learning when some or all nodes of the graph are latent variables,
  accommodating non-Gaussian manifest variables such as binary and categorical data in the measurement model. To overcome the computational burden of high-dimensional integration in standard marginal likelihoods, we develop DAG learning via Composite-Likelihood-based screening and Iterative Pruning (DAG-CLIP), a two-step composite-likelihood-based learning algorithm. This algorithm first solves a smooth acyclicity-constrained optimization problem to screen out misspecified DAG structures, and subsequently performs BIC-guided composite-likelihood backward deletion within the Markov equivalence class (MEC) space to identify a sparse DAG. We establish the statistical consistency of the algorithm in recovering the MEC of the true DAG, and demonstrate its effectiveness through extensive simulation studies and a real-world application to large-scale educational survey data.
\end{abstract}
 
\begin{keywords}
  Latent variables, causal discovery, Markov equivalence class, pairwise likelihood, composite-likelihood, NOTEARS
\end{keywords}
\newpage

\section{Introduction}\label{sec:intro}

\subsection{Background }\label{subsec:intro_background}

Directed graphical models characterize dependencies among variables through a directed graph. A prominent subclass uses directed acyclic graphs (DAGs), which are well suited to representing causal relationships \citep[see Chapter 2,][]{pearl2009causality}. Consequently, there is great interest in learning DAGs to uncover underlying causal mechanisms. This task, known as causal discovery, has received widespread attention across machine learning, statistics, and data science \citep[see, e.g.,][]{SpirtesGlymourScheines2000,chickering2002optimal,kalisch2007pc,ZhengEtAl2018,yuan2019constrained,zhao2022nonGaussian,wang2024peeling}.

A key limitation of most existing DAG learning algorithms is their strict requirement that all variables of interest (i.e., DAG nodes) must be directly observed. However, in many applications, especially those in social and behavioral research, many variables of substantive interest are not directly observable.
Examples include personality traits, cognitive abilities, mental health status, and political attitudes. These constructs are typically formulated as latent variables and defined through a statistical measurement model—a generative model that specifies the relationship between the latent factors and their manifest indicators. Depending on the data type of the manifest indicators, this measurement model may take different forms, such as a linear factor model \citep[Chapter 7,][]{bollen1989structural}, including its underlying-variable formulation for binary and ordinal indicators \citep{Muthen1984}, or an item response theory (IRT) model \citep{ChenLiLiuYing2025}.

In recent years, several methods have been proposed for DAG learning in which the nodes are continuous latent variables. Most of them learn the measurement model from data by first identifying groups of manifest variables that each measure a single latent variable and then learning the structure among the identified latent variables \citep{SilvaScheinesGlymourSpirtes2006,kummerfeld2016causal,cai2019triad,xie2020generalized,huang2022latent}, or by score-based search \citep{NgEtAl2024SALAD,DongEtAl2026LGES}. {These approaches mainly consider continuous manifest variables, although recent rank-based methods also accommodate discrete variables \citep{DongEtAl2025MPRT}. 
{In these approaches, the measurement structure between latent variables and their observed indicators is learned from data rather than specified in advance.}
}
Methods that take the measurement model as given either allow a single indicator per latent variable \citep{LuoMoffaKuipers2021,Castelletti2024} or, as in the MIMBuild algorithm \citep{SpirtesGlymourScheines2000,SilvaScheinesGlymourSpirtes2006}, assume a linear factor model with continuous manifest variables.
{Our focus is on a prespecified (confirmatory) measurement structure that allows each latent variable to be measured by multiple categorical or mixed-type indicators, a common pattern in social and behavioral research.} {Moreover, the computational cost is high when the measurement model is non-Gaussian, as the likelihood involves an integral over the latent variables that has no closed form.} %
We refer the readers to Section~\ref{sec:related} for further discussion.

\subsection{Contributions}\label{subsec:intro_contr}

This paper addresses the gaps mentioned above in existing DAG learning algorithms when latent variables are present. The contribution is fourfold:

\begin{enumerate}
  \item We propose a general modeling framework for DAG learning,  when some or all nodes of the graph are latent variables. Specifically, we assume that the latent variables follow a multivariate normal distribution with a covariance matrix induced by a DAG. Meanwhile, we impose minimal restrictions on the measurement model, thereby allowing virtually any specification, subject to standard smoothness and identifiability requirements for the unknown model parameters.

  \item We develop a computationally efficient {two-step} algorithm {with statistical guarantees} to learn the latent DAG. The first step screens the space of DAGs using a sparsity-inducing penalty and returns a DAG compatible with the true latent distribution with probability tending to one. The second step removes the spurious edges by a backward search over Markov equivalence classes (MECs). The two steps rely on four technical tools. First, we employ a pairwise composite likelihood \citep{Lindsay1988,VarinReidFirth2011,KatsikatsouEtAl2012,KatsikatsouMoustaki2016} as part of the objective function {in both steps}.
        A challenge brought by the extended measurement model is that
        the likelihood function involves an integral over all latent variables, which admits a closed-form solution only for the Gaussian linear factor model. When the number of latent variables is moderately large, evaluating the likelihood function repeatedly during graph search becomes computationally prohibitive. The pairwise likelihood addresses this challenge. In particular, each term of the pairwise likelihood involves an integral of dimension at most two, regardless of the number of latent variables. It therefore avoids the high-dimensional integration required by the marginal likelihood.
        {Second, the screening step imposes sparsity through a truncated $\ell_1$ penalty, which is a continuous surrogate for the $\ell_0$ penalty. At the global solution of the penalized problem, this penalty yields consistent selection of the sparse structure \citep{shen2012likelihood,yuan2019constrained}, a property that the $\ell_1$ penalty 
        does not guarantee \citep{vandegeer2013l0}. The penalized problem is nonconvex and is solved by difference-of-convex programming \citep{lethi2018dc} through a sequence of weighted $\ell_1$ problems. 
        Third, the smooth acyclicity constraint of NOTEARS \citep{ZhengEtAl2018}, originally developed for fully observed variables, turns the combinatorial search over DAGs into a continuous optimization problem in which all candidate edges are updated jointly. 
        Fourth, a composite likelihood Bayesian information criterion (CLBIC; \citealp{GaoSong2010}) constructed from the same pairwise likelihood guides the pruning step. Based on these tools, we name the proposed algorithm DAG-CLIP (DAG learning via Composite-Likelihood-based screening and Iterative Pruning).}

        Both screening and pruning are important for effective DAG learning. While it is possible to learn a DAG using only the pruning step by starting from a fully connected DAG, as in \cite{DongEtAl2026LGES}, this approach has two major drawbacks. First, the computational burden of iterative pruning can be prohibitive because a dense DAG generates many search candidates per iteration. Second, the expanded search space increases the statistical error introduced by multiple comparisons when using the CLBIC. We can also learn a DAG by running only the screening step. In fact, this approach can be shown to consistently recover the MEC of the 
        true sparse DAG. {However, this guarantee requires a sufficiently accurate approximation to the global maximum of a highly nonconvex objective, which may be difficult to obtain in practice. Our simulations in Section~\ref{sec:simulation} show that screening alone often retains spurious edges, and they are only eliminated in the pruning steps. The pruning step permits consistent selection under a weaker numerical accuracy condition on the screening solution.} This learning strategy of integrating these two steps is modular, meaning it can be seamlessly combined with other suitable objective functions for DAG learning, such as the marginal likelihood under Gaussian DAG models.

  \item We show that the proposed algorithm can consistently recover the MEC of the true DAG {under a numerical accuracy condition on the screening solution that is weaker than the condition used to establish consistency of screening alone.}
        This result aligns with standard statistical consistency results for DAG learning from observational data without latent variables \citep[e.g.,][]{chickering2002optimal,kalisch2007pc,nandy2018highdimensional}. That is, without additional assumptions or experimental interventions, the underlying DAG is identifiable only up to its MEC.

  \item Finally, we empirically evaluate the proposed method through extensive simulations and an application to large-scale educational survey data.
        Our simulations compare DAG-CLIP with {its screening step alone and with} two-step methods that first fit the measurement model and then apply the PC algorithm \citep{SpirtesGlymourScheines2000} or greedy equivalence search \citep{chickering2002optimal} to estimated factor scores or latent correlations, demonstrating that the proposed approach substantially outperforms these competitors across diverse settings. Furthermore, we apply the method to the Australian Grade 8 public-use data from the 2023 Trends in International Mathematics and Science Study (TIMSS) \citep{vonDavierFishbeinKennedy2024} to learn a DAG involving 14 latent variables and 2 observed variables, where the latent constructs are measured by 69 manifest indicators comprising both binary and ordinal types. The resulting structure yields highly interpretable insights. {In addition, with fewer than half of the possible adjacencies, it predicts held-out responses nearly as well as the saturated model.}
\end{enumerate}

\subsection{Related Literature}\label{sec:related}

This work connects to several key areas in machine learning and statistics. We organize the review around three topics, namely DAG learning of observed variables, DAG learning in the presence of latent variables, and composite likelihood.

When all relevant variables are directly observed, DAG learning methods typically focus on recovering the DAG up to its MEC. They may be classified into three broad groups. First, constraint-based methods decide adjacencies and orientations through hypothesis tests of conditional independence. A representative method is the PC algorithm \citep{SpirtesGlymourScheines2000,kalisch2007pc}, while other methods have also been developed in this direction, such as conservative PC \citep{ramsey2006adjacency}, rank PC for nonparanormal data \citep{harris2013pc}, and the order-independent PC-stable \citep{colombo2014order}.
Second, score-based methods select a graph by optimizing a model-selection criterion over graphs or equivalence classes. Examples include greedy equivalence search \citep{chickering2002optimal}, which builds on the classical selection consistency of the BIC \citep{haughton1988choice}, and {penalized likelihood formulations with the $\ell_0$ penalty or its nonconvex surrogates \citep{vandegeer2013l0,aragam2015concave,yuan2019constrained}. In particular, \citet{yuan2019constrained} reconstruct a Gaussian DAG by a constrained likelihood with truncated $\ell_1$ constraints. {Because these optimization problems are nonconvex, obtaining the solutions required by the statistical guarantees can be difficult in practice.}} Finally, continuous-optimization methods {express acyclicity through smooth constraints or penalties} and update all candidate edges together through gradient-based optimization \citep{ZhengEtAl2018,yu2019dag,ng2020golem,bello2022dagma}. {These methods commonly use an $\ell_1$-type penalty to impose sparsity.} {The proposed method can be seen as a hybrid of methods from the last two groups. Its screening step uses continuous optimization with a truncated $\ell_1$ penalty, and its pruning step uses a score-based search.}

Methods have also been developed to distinguish the true DAG from the rest of its MEC by incorporating additional assumptions or prior information. Examples include linear models with non-Gaussian errors \citep{shimizu2006linear,zhao2022nonGaussian}, Gaussian models with equal error variances \citep{peters2014identifiability,chen2019causal}, parametric ordinal models \citep{ni2022ordinal,ni2025causal}, and methods that use interventional data or instrumental variables \citep{ellis2008learning,peng2020reconstruction,li2023inference,chen2024discovery,wang2024peeling,taeb2024latent}.

As mentioned earlier, DAG learning in the presence of latent variables has received increasing attention in recent years. Methods can be divided into two groups based on the role of the latent variables. In the first group, the latent variables are unobserved confounders, and the graph nodes remain the observed variables. \cite{SpirtesGlymourScheines2000,ColomboMaathuisKalischRichardson2012}
proposed the FCI algorithm and its variants, which learn an ancestral graph over the observed variables in the presence of unobserved confounders, and \citet{ChenDashGao2021} developed an exact score-based method of this type using integer programming.

In the second group, the latent variables are treated as nodes in the DAG, and methods differ on whether the measurement structure, i.e., the relationships between the latent variables and the observed indicators, is learned from data or specified in advance. When the measurement structure is learned, the assignment of observed variables to latent variables is estimated together with the graph. Most methods first identify groups of observed variables that each measure a single latent variable by testing constraints on the observed covariance matrix, and then learn the structure among the latent variables \citep{SilvaScheinesGlymourSpirtes2006,kummerfeld2016causal,cai2019triad,xie2020generalized,huang2022latent}. Score-based methods have also been developed for this problem \citep{NgEtAl2024SALAD,DongEtAl2026LGES}. {In particular, \citet{DongEtAl2026LGES} propose a latent variable greedy equivalence search that starts from a fully connected graph and removes edges by a backward search.} {\citet{DongEtAl2025MPRT} develop a mixed data permutation-based rank test (MPRT) for covariance-rank constraints with discretized measurements and demonstrate its use in latent causal discovery.
In these methods, latent variables are defined by the estimated grouping of indicators rather than by substantive theory, so their interpretation can sometimes be unclear.
}

When the measurement model is specified in advance, the method learns only the structure among the latent variables. This is the setting of the proposed method.
In the simplest case, each latent variable is measured by a single observed variable, as in latent Gaussian DAG models for ordinal and mixed-type data \citep{LuoMoffaKuipers2021,Castelletti2024}.
These models are not flexible enough to include latent variables each measured by multiple indicators. These methods involve optimization based on marginal likelihood functions that involve high-dimensional integrals and tackle the optimization using either
Monte Carlo Expectation-Maximization or full Markov chain Monte Carlo algorithms, which are computationally intensive.
When each latent variable is measured by several indicators with a given measurement structure, the MIMBuild algorithm \citep{SpirtesGlymourScheines2000,SilvaScheinesGlymourSpirtes2006} learns the DAG of the latent variables either by applying the PC algorithm to conditional independence tests derived from rank constraints on the covariance matrix of the indicators, or by greedy equivalence search under a Gaussian likelihood fitted by an EM algorithm.  It is developed for linear models with continuous indicators and does not accommodate categorical responses.  DAG learning with discrete latent variables has also been studied, as in \cite{GuXu2019}, \cite{liu2022measurement}, \cite{LeeGu2024}, and \cite{zhang2026discrete}.

Finally, we review the literature on composite likelihood, which appears in the objective function of the proposed method. Composite likelihood is an inference function formed by multiplying a collection of component likelihoods \citep{Lindsay1988,VarinReidFirth2011}. In this work, we employ a variant known as pairwise likelihood, constructed from bivariate margins. This approach has been widely used to estimate high-dimensional latent variable models due to its computational advantages over the standard marginal likelihood \citep{KatsikatsouEtAl2012, KatsikatsouMoustaki2016, Alfonzetti2024, Alfonzetti2025composite}. {However, to the best of our knowledge, no existing work uses the pairwise likelihood or other composite likelihood to score and select DAG structures.} Model selection under a composite likelihood can be performed using CLBIC \citep{GaoSong2010}, a criterion whose penalty function uses an estimated model degree of freedom—rather than simply the number of unknown parameters—to quantify model complexity. The proposed algorithm leverages this criterion during the iterative pruning step.

The rest of the paper is organized as follows. {In Section~\ref{sec:model_and_problem_settings}, we introduce the modeling framework and discuss the identifiability of the DAG. Section~\ref{sec:alg} presents {the technical tools} and the DAG-CLIP algorithm. Section~\ref{sec:iden} establishes the consistency of DAG-CLIP. Sections~\ref{sec:simulation} and \ref{sec:real} present simulation studies and an application to the TIMSS 2023 data. Section~\ref{sec:dis} concludes with a discussion of the limitations and future directions. Proofs of the theoretical results, computational details, and additional details of the simulation study and the application are provided in the supplementary material, which is available from the authors upon request.}

\section{Model Framework and Identifiability}\label{sec:model_and_problem_settings}

\subsection{Model Framework}\label{subsec:model}

\paragraph{Model.}
Suppose that we are interested in learning the DAG of $m$ variables, denoted by $\boldsymbol\eta=(\eta_{1},\ldots,\eta_{m})^\top$. Without loss of generality, we assume that they are all latent variables. As we discuss later, our framework also allows certain variables to be observed.

We assume that $\boldsymbol\eta$ follows an acyclic Gaussian linear structural equation model,
\begin{equation}\label{eq:lsem}
  \boldsymbol\eta = B^\top\boldsymbol\eta + \boldsymbol\zeta,
  \qquad
  \boldsymbol\zeta\sim\mathcal N(\mathbf 0,\Omega),
  \qquad
  \Omega=\diag(\omega_1,\ldots,\omega_m)\succ0,
\end{equation}
where $B=(\beta_{jk})_{m\times m}$ is the coefficient matrix and $\beta_{jk}$ is the effect of $\eta_j$ on $\eta_k$. The support of $B$ encodes a DAG: the directed graph $G=(V,E)$ with $V=\{1,\ldots,m\}$ and $E=\{(j,k):\beta_{jk}\neq0\}$ has no directed cycles. {We further assume that the true DAG is sparse, as is common in DAG learning \citep{kalisch2007pc,ZhengEtAl2018}.} Model \eqref{eq:lsem} is the latent variable model of linear structural equation modeling, as presented in \citet[Chapter 8]{bollen1989structural} and \citet{BollenPearl2013}, without exogenous variables; see \citet[Chapter 5]{pearl2009causality} for its causal interpretation. Since $G$ is acyclic, $I-B$ is nonsingular, and \eqref{eq:lsem} implies
\begin{equation}\label{eq:lsem_cov}
  \boldsymbol\eta\sim\mathcal N(\mathbf 0,\Phi),
  \qquad
  \Phi=(I-B^\top)^{-1}\Omega(I-B)^{-1}.
\end{equation}
The zero mean fixes the location of the latent variables. To fix their scale, we constrain each latent variable to have unit variance, that is, $\Phi_{ii}=1$ for $i=1,\ldots,m$.

Further suppose that $\boldsymbol\eta$ are measured by $p$ indicators, denoted by $\mathbf Y=(Y_{1},\ldots,Y_{p})^\top$. A measurement model specifies the conditional distribution of $\mathbf Y$ given $\boldsymbol\eta$. We make the following assumptions:
\begin{enumerate}
  \item Local independence assumption: $Y_1$, ..., $Y_p$ are conditionally independent given $\boldsymbol\eta$.
  \item Unidimensional measurement and parametric conditional density: Each variable $Y_j$ is only associated with a single latent variable, $\eta_{k_j}$ and the conditional distribution of $Y_j$ given ${\eta}_{k_j}$ has a parametric probability density/mass function
        $f_j(y \mid \eta_{k_j}; \boldsymbol\alpha)$, where $\boldsymbol\alpha \in \mathcal A$ is a generic notation containing the unknown parameters in these functions and $\mathcal A$ denotes the parameter space for $\boldsymbol{\alpha}$ to incorporate possible constraints on $\boldsymbol\alpha$.
\end{enumerate}

We provide some examples below.

\begin{example}[Graded response model]\label{ex:grm}
  Suppose that $Y_j$ is an ordinal variable with $C_j\ge2$ ordered categories, coded $y\in\{0,1,\ldots,C_j-1\}$, that measures $\eta_{k_j}$. The graded response model \citep{Samejima1969} in its probit form specifies
  \begin{equation}\label{eq:grm}
    Y_j=c
    \quad\Longleftrightarrow\quad
    \tau_{j,c}<\lambda_j\eta_{k_j}+e_j\le\tau_{j,c+1},
    \qquad e_j\sim\mathcal N(0,1),
  \end{equation}
  where $e_j$ is independent of $\boldsymbol\eta$, $\lambda_j$ is the loading, and $-\infty=\tau_{j,0}<\tau_{j,1}<\cdots<\tau_{j,C_j-1}<\tau_{j,C_j}=\infty$ are thresholds \citep[see also Chapter 5 of][]{BartholomewKnottMoustaki2011}. Hence
  \begin{equation}\label{eq:grm_pmf}
    f_j(y\mid\eta_{k_j};\boldsymbol\alpha)
    =\int_{\tau_{j,y}-\lambda_j\eta_{k_j}}^{\tau_{j,y+1}-\lambda_j\eta_{k_j}}\varphi(t)\,\mathrm dt,
    \qquad y\in\{0,1,\ldots,C_j-1\},
  \end{equation}
  where $\varphi$ is the standard normal density. In this case, $\boldsymbol\alpha$ consists of $(\lambda_j,\tau_{j,1},\ldots,\tau_{j,C_j-1})$ for $j=1,\ldots,p$, and the parameter space $\mathcal A$ imposes an ordering constraint $\tau_{j,1}<\cdots<\tau_{j,C_j-1}$ for each $j$. The unit variance of $e_j$ fixes the scale of the underlying response. For a binary item, $C_j=2$ and there is a single threshold $\tau_{j,1}$.
\end{example}

\begin{example}[Single-indicator threshold model]\label{ex:single_threshold}
  {Suppose that every latent variable is measured by a single ordinal variable, so that $p=m$ and $Y_k$ measures $\eta_k$ for $k=1,\ldots,m$.
    Let $Y_k$ have $C_k\ge2$ ordered categories. With one indicator, the underlying response is taken to be $\eta_k$ itself:}
  \begin{equation}\label{eq:single_threshold}
    Y_k=c
    \quad\Longleftrightarrow\quad
    \tau_{k,c}<\eta_k\le\tau_{k,c+1},
    \qquad c=0,1,\ldots,C_k-1,
  \end{equation}
  where $-\infty=\tau_{k,0}<\tau_{k,1}<\cdots<\tau_{k,C_k-1}<\tau_{k,C_k}=\infty$ are thresholds. Hence $f_k(y\mid\eta_k;\boldsymbol\alpha)$ equals one if $\tau_{k,y}<\eta_k\le\tau_{k,y+1}$ and zero otherwise, and the parameters of item $k$ in $\boldsymbol\alpha$ are the thresholds $\tau_{k,1},\ldots,\tau_{k,C_k-1}$, subject to the ordering in $\mathcal A$. It is a special case of the graded response model in Example~\ref{ex:grm} when all the loading parameters are constrained to be 1 and the error terms are set to be zero.
  The same model
  has been considered in \citet{LuoMoffaKuipers2021} and  \citet{Castelletti2024}.
\end{example}

\begin{example}[Linear factor model]\label{ex:linear_factor}
  Suppose that $Y_j$ is a continuous variable that measures $\eta_{k_j}$. The linear factor model \citep[Chapter 7,][]{bollen1989structural} specifies
  \begin{equation}\label{eq:linear_factor}
    Y_j=d_j+\lambda_j\eta_{k_j}+\varepsilon_j,
    \qquad
    \varepsilon_j\sim\mathcal N(0,\psi_j),
  \end{equation}
  where $\varepsilon_j$ is independent of $\boldsymbol\eta$, $d_j$ is an intercept, $\lambda_j$ is the loading, and $\psi_j>0$ is the residual variance. Hence
  \begin{equation}\label{eq:linear_factor_pdf}
    f_j(y\mid\eta_{k_j};\boldsymbol\alpha)
    =\psi_j^{-1/2}\,\varphi\{\psi_j^{-1/2}(y-d_j-\lambda_j\eta_{k_j})\},
    \qquad y\in\mathbb R,
  \end{equation}
  and the parameters of item $j$ in $\boldsymbol\alpha$ are $(d_j,\lambda_j,\psi_j)$, subject to $\psi_j>0$ in $\mathcal A$. When every latent variable is measured in this way, $\mathbf Y$ is multivariate normal, and the model is the confirmatory factor analysis model with simple structure, for which the marginal likelihood has a closed form.
\end{example}

\begin{example}[Model for an observed variable]\label{ex:observed}
  Suppose that node $k$ of the DAG is a directly observed continuous variable $Y_j$. The framework handles this case by taking $\eta_k$ to be $Y_j$ up to location and scale, without measurement error,
  \begin{equation}\label{eq:observed}
    Y_j=d_j+a_j\eta_k,
  \end{equation}
  where $a_j>0$ in $\mathcal A$. Then $f_j(\cdot\mid\eta_k;\boldsymbol\alpha)$ becomes a Dirac measure, and the parameters of item $j$ in $\boldsymbol\alpha$ are $(d_j,a_j)$.
\end{example}

We note that the current model framework assumes that each manifest variable measures only one latent variable, which is probably the most common setting in practice. We make this assumption to simplify the notation and subsequent discussion. We note that it is easy to extend the modeling framework to allow $Y_j$ to depend on multiple $\eta_k$s; {see Section~\ref{sec:dis} for further discussion.}

\subsection{Marginal Likelihood and Identifiability}

Let $\mathbf Y_i=(Y_{i1},\ldots,Y_{ip})^\top$, $i=1,\ldots,n$, be independent copies of $\mathbf Y$, and let $\boldsymbol\theta=(\boldsymbol\alpha,B,\Omega)$ collect the parameters of the measurement and structural models{, with true value $\boldsymbol\theta^*=(\boldsymbol\alpha^*,B^*,\Omega^*)$ under which the data are generated.} We denote $\Phi^* = (I-(B^*)^\top)^{-1}\Omega^*(I-B^*)^{-1}$, which  satisfies $\Phi^*_{ii}=1$ for $i=1,\ldots,m$. Further, let $G^*$ be the DAG encoded by the support of $B^*$. {As mentioned earlier, $G^*$ is assumed to be sparse.}

Under the local independence assumption, the marginal density of $\mathbf Y$ is
\begin{equation}\label{eq:marginal_density}
  p(\mathbf y;\boldsymbol\theta)
  =\int_{\mathbb R^m}\left(\prod_{j=1}^p f_j(y_j\mid\eta_{k_j};\boldsymbol\alpha)\,
  \right)\varphi_m(\boldsymbol\eta;\mathbf 0,\Phi)\,\mathrm d\boldsymbol\eta,
\end{equation}
where $\varphi_m(\cdot\,;\mathbf 0,\Phi)$ is the $m$-variate normal density with mean $\mathbf 0$ and covariance matrix $\Phi$ given in \eqref{eq:lsem_cov}. The marginal log-likelihood is $\ell_n(\boldsymbol\theta)=\sum_{i=1}^n\log p(\mathbf Y_i;\boldsymbol\theta)$.

The integral in \eqref{eq:marginal_density} is of dimension $m$. It has a closed form when the manifest variables either follow the linear factor model in Example \ref{ex:linear_factor} or are observed as in Example
\ref{ex:observed}, in which case $\mathbf Y$ is multivariate normal. For categorical manifest variables, as in Examples~\ref{ex:grm} and \ref{ex:single_threshold}, it has no closed form and {has to be approximated by computational methods, such as numerical integration via quadrature points, where the number of points grows exponentially in $m$, or Monte Carlo integration, which requires a large number of draws for sufficient accuracy}. Structure learning compounds this cost, since the likelihood is evaluated repeatedly over candidate graphs and parameter values. Direct maximization of $\ell_n$ is therefore computationally prohibitive when $m$ is moderately large. This is the computational challenge that the pairwise likelihood introduced below addresses.

Since \eqref{eq:marginal_density} depends on $(B,\Omega)$ only through $\Phi$, we write $p(\mathbf y;\boldsymbol\alpha,\Phi)$. Let $\mathcal S_m$ denote the set of $m\times m$ positive definite correlation matrices, to which $\Phi$ belongs by \eqref{eq:lsem_cov}. The sign of each latent variable is fixed by requiring one of its loadings to be positive, which we include in $\mathcal A$. We assume throughout that the measurement model is identifiable in the following sense.

\begin{assumption}[Identifiability of the measurement model]\label{ass:measurement_id}
  {For any $(\boldsymbol\alpha,\Phi)\in\mathcal A\times\mathcal S_m$, $p(\mathbf y;\boldsymbol\alpha,\Phi)=p(\mathbf y;\boldsymbol\alpha^*,\Phi^*)$ for all $\mathbf y$ implies $(\boldsymbol\alpha,\Phi)=(\boldsymbol\alpha^*,\Phi^*)$.}
\end{assumption}

Assumption~\ref{ass:measurement_id} holds under mild conditions on the measurement design. {For the graded response model in Example~\ref{ex:grm}, identifiability conditions are given by \citet{WuEstabrook2016}.} For the single-indicator threshold model in Example~\ref{ex:single_threshold}, Assumption~\ref{ass:measurement_id} holds without further conditions, since the thresholds and the tetrachoric or polychoric correlations $\Phi_{jk}$ are identified from the univariate and bivariate margins of $\mathbf Y$ \citep{Pearson1900,Olsson1979,LuoMoffaKuipers2021}.
For the linear factor model in Example~\ref{ex:linear_factor}, it suffices that each latent variable has three indicators, or two if it is correlated with another latent variable (\citealp{anderson1956statistical}, Chapter 7, \citealp{bollen1989structural}).

Under Assumption~\ref{ass:measurement_id}, $\Phi^*$ is identified from the distribution of $\mathbf Y$. Identifying $G^*$ then reduces to the well-understood problem of identifying a Gaussian DAG model from $\boldsymbol\eta\sim\mathcal N(\mathbf 0,\Phi^*)$. First, the true DAG is typically not identifiable from this normal distribution, or equivalently, the true covariance matrix $\Phi^*$. {$G^*$ belongs to a Markov equivalence class (MEC), the set of DAGs that encode the same conditional independence relationships as $G^*$. We say that a DAG induces a correlation matrix $\Phi$ if $\Phi=(I-B^\top)^{-1}\Omega(I-B)^{-1}$ for some $B$ whose support is contained in its edge set and some diagonal $\Omega\succ0$. Every DAG in this class induces $\Phi^*$.} This Markov equivalence class is fully characterized by the so-called completed partially directed acyclic graph (CPDAG), a graph with both directed and undirected edges constructed from $G^*$ that has a one-to-one correspondence with the Markov equivalence class \citep{andersson1997characterization,meek1995causal,chickering2002equivalence}. Second, {all DAGs in the class have the same number of edges as $G^*$, and} with an additional faithfulness condition (see, e.g., Section 2.3.3, \citealp{SpirtesGlymourScheines2000}, and Assumption~\ref{ass:faithfulness} in Section~\ref{sec:iden}), {every DAG outside the class that induces $\Phi^*$ has strictly more edges} \citep[Theorem 4,][]{chickering2002optimal}. {The MEC of $G^*$ is therefore the unique sparsest class of DAGs that induces $\Phi^*$.} In the rest of the paper, we focus on estimating the MEC of $G^*$.

\section{Proposed Method}\label{sec:alg}

{The proposed DAG-CLIP has two steps. The first step performs a screening in the space of DAGs with sufficient sparsity and returns a DAG and 
its MEC. As shown in Section~\ref{sec:iden}, with high probability, this DAG induces $\Phi^*$ but may contain spurious edges. The second step removes spurious edges by searching backwards over the space of MECs. 
These two steps rely on {four}
technical tools, including (1) a pairwise likelihood that is involved in the objective functions in both steps, (2) a truncated $\ell_1$ penalty, which is used in the first step to impose sparsity on the search space, {(3) a smooth acyclicity constraint, which 
is incorporated into the objective function in the first step,} and {(4)} a composite likelihood Bayesian information criterion that guides the pruning step. We introduce these tools in Section~\ref{subsec:pairwise} and then describe the two steps in Sections~\ref{subsec:screening} and \ref{subsec:pruning}, respectively.}

\subsection{Technical Tools}\label{subsec:pairwise}

\paragraph{Pairwise likelihood.}
Recall that $\mathbf Y_1,\ldots,\mathbf Y_n$ are independent copies of $\mathbf Y$. The pairwise likelihood replaces the marginal density \eqref{eq:marginal_density} by the bivariate marginal densities of all pairs of manifest variables. For $1\le j<l\le p$ with $k_j\ne k_l$, the bivariate marginal density of $(Y_j,Y_l)$ is
\begin{equation}\label{eq:pair_margin}
  p_{jl}(y_j,y_l;\boldsymbol\alpha,\Phi)
  =\int_{\mathbb R^2} f_j(y_j\mid\eta_{k_j};\boldsymbol\alpha)\,f_l(y_l\mid\eta_{k_l};\boldsymbol\alpha)\,
  \varphi_2(\eta_{k_j},\eta_{k_l};\Phi_{k_jk_l})\,\mathrm d\eta_{k_j} \mathrm d\eta_{k_l},
\end{equation}
where $\varphi_2(\cdot,\cdot\,;\rho)$ is the bivariate standard normal density with correlation $\rho$. When $k_j=k_l$, the two variables measure the same latent variable, and \eqref{eq:pair_margin} is reduced to a one-dimensional integral over $\eta_{k_j}$ with the standard normal density. In either case, the integral has dimension at most two. We use $\boldsymbol\theta=(\boldsymbol\alpha,B,\Omega)$ as the generic notation for all the unknown parameters.
The pairwise log-likelihood is
\begin{equation}\label{eq:pair_loglik}\noeqref{eq:pair_loglik}
  \ell_n^{\mathrm{pair}}(\boldsymbol\theta)
  =\sum_{i=1}^n cl_i(\boldsymbol\theta),
  \qquad \mbox{where~}
  cl_i(\boldsymbol\theta)
  =\sum_{1\le j<l\le p}\log p_{jl}(Y_{ij},Y_{il};\boldsymbol\alpha,\Phi),
\end{equation}
with $\Phi$ being a function of $B$ and $\Omega$ given by \eqref{eq:lsem_cov}.

To replace the standard marginal likelihood with the pairwise likelihood, we need to ensure that the measurement model remains identifiable under the pairwise likelihood. Thus, we make the following Assumption~\ref{ass:pairwise_id}, which extends Assumption~\ref{ass:measurement_id} to the pairwise likelihood.
  {For the models in Examples~\ref{ex:grm}--\ref{ex:observed}, Assumption~\ref{ass:measurement_id} implies Assumption~\ref{ass:pairwise_id}; see the supplementary material for a proof.}

\begin{assumption}[Identifiability from bivariate margins]\label{ass:pairwise_id}
  {For $(\boldsymbol\alpha,\Phi)\in\mathcal A\times\mathcal S_m$, if
    \[
      p_{jl}(y_j,y_l;\boldsymbol\alpha,\Phi)=p_{jl}(y_j,y_l;\boldsymbol\alpha^*,\Phi^*)
    \]
    for all $1\le j<l\le p$ and all $(y_j,y_l)$, then $(\boldsymbol\alpha,\Phi)=(\boldsymbol\alpha^*,\Phi^*)$.}
\end{assumption}

When a DAG $G$ is given, we use $\Theta_G$ to denote the parameter space for $\boldsymbol\theta$, which encodes the constraint that
the support of $B$ is contained in the edge set of $G$, as well as the identification constraints in Section~\ref{subsec:model}. In this case, the
maximum pairwise likelihood estimator of $\boldsymbol\theta$ subject to DAG $G$
is $\widehat{\boldsymbol\theta}_G=\argmax_{\boldsymbol\theta\in\Theta_G}\ell_n^{\mathrm{pair}}(\boldsymbol\theta)$. Based on the asymptotic theory for composite likelihood \citep[see, e.g., ][]{VarinReidFirth2011}, when $G$ is in the MEC of $G^*$ and under suitable regularity conditions, $\widehat{\boldsymbol\theta}_G$ is consistent and asymptotically normal. Let $\Theta$ denote the union of $\Theta_G$ over all DAGs $G$ with nodes $\{1,\ldots,m\}$. Then $\boldsymbol\theta\in\Theta$ if and only if the support of $B$ is acyclic, subject to the rest of the identification constraints. {The screening step maximizes a penalized version of $\ell_n^{\mathrm{pair}}$ over $\Theta$, and the pruning step compares CPDAGs by the CLBIC computed from $\ell_n^{\mathrm{pair}}$.}

\paragraph{Truncated $\ell_1$ penalty.}
{The screening step searches over DAGs by penalized continuous optimization. It relies on a sparsity penalty that is used to screen out dense DAGs, as the true DAG is assumed to be sparse. 
More specifically, we use the truncated $\ell_1$ function of \citet{shen2012likelihood} as a continuous surrogate of the $\ell_0$ norm of $B$.}

{For $\tau>0$, let}
\begin{equation}\label{eq:tlp}
  J_\tau(x)=\min\left(\frac{|x|}{\tau},1\right),\qquad x\in\mathbb R.
\end{equation}
{It equals one when $|x|\ge\tau$ and is linear below. Hence $\sum_{j\ne k}J_\tau(\beta_{jk})$ is continuous in $B$ and equals the number of nonzero entries of $B$ once $\tau$ is below the smallest nonzero $|\beta_{jk}|$. Nonconvex penalties and constraints of this type have been used for DAG learning over observed Gaussian variables \citep{aragam2015concave,yuan2019constrained}.}
{The truncated $\ell_1$ function is a difference of two convex functions, $J_\tau(x)=|x|/\tau-\max(|x|/\tau-1,0)$. A problem penalized by $J_\tau$ can therefore be solved by difference-of-convex iterations \citep{shen2012likelihood}.}

\paragraph{Smooth acyclicity constraint.}
The acyclicity constraint on the support of $B$ is combinatorial. {Following the NOTEARS approach of \citet{ZhengEtAl2018}, we express it as the smooth equality constraint $h(B)=0$ where}
\begin{equation}\label{eq:acyclicity_new}
  h(B)=\tr\{\exp(B\circ B)\}-m.
\end{equation}

Here, the notation $\circ$ denotes the Hadamard product and $\exp(\cdot)$ the matrix exponential. The function $h$ is nonnegative and equals zero if and only if the support of $B$ is acyclic. {The resulting continuous formulation allows candidate edge coefficients to be updated jointly.}
{Under this constraint, each weighted $\ell_1$ problem of the difference-of-convex iterations has the form of the NOTEARS problem and is solved by the augmented Lagrangian method.}
{Alternative smooth characterizations of acyclicity include the log-determinant constraint used in \cite{bello2022dagma}.}
 
\paragraph{CLBIC.} Model selection under the pairwise likelihood can be performed
based on the composite likelihood Bayesian information criterion (CLBIC; \citealp{GaoSong2010}). For a DAG $G$ and given the sample size $n$, define
\begin{equation}\label{eq:clbic_def}
  \operatorname{CLBIC}_n(G)
  =-2\,\ell_n^{\mathrm{pair}}(\widehat{\boldsymbol\theta}_G)
  +\widehat d_{\mathrm{eff}}(G)\log n,
  \qquad
  \widehat d_{\mathrm{eff}}(G)=\tr(\widehat J_G\widehat H_G^{-1}),
\end{equation}
where $\widehat H_G=-n^{-1}\sum_{i=1}^n\nabla^2cl_i(\widehat{\boldsymbol\theta}_G)$ and $\widehat J_G=n^{-1}\sum_{i=1}^n\nabla cl_i(\widehat{\boldsymbol\theta}_G)\nabla cl_i(\widehat{\boldsymbol\theta}_G)^\top$ are estimates of the
sensitivity and variability matrices, respectively.
Here $\nabla cl_i$ and $\nabla^2cl_i$ denote the gradient and Hessian of $cl_i$ with respect to the free parameters in $\Theta_G$. The sensitivity and variability matrices are $H(\boldsymbol\theta)=-\mathbb E\{\nabla^2cl_i(\boldsymbol\theta)\}$ and $J(\boldsymbol\theta)=\operatorname{Var}\{\nabla cl_i(\boldsymbol\theta)\}$.
The penalty replaces the number of free parameters in the ordinary BIC by an estimate of the effective number of parameters, which accounts for the dependence among the terms of the pairwise likelihood. {In DAG-CLIP, the CLBIC scores the candidate CPDAGs in the pruning step and selects the truncation level of the penalty in the screening step.}

\subsection{Step 1: Screening}\label{subsec:screening}

We define the penalized pairwise log-likelihood as
\begin{equation}\label{eq:lpen}
  \ell_n^{\mathrm{pen}}(\boldsymbol\theta)
  =\ell_n^{\mathrm{pair}}(\boldsymbol\theta)
  -n\lambda_n\sum_{j\ne k}J_{\tau_n}(\beta_{jk}),
\end{equation}
where $\lambda_n>0$ and $\tau_n>0$ are tuning parameters and $J_{\tau_n}$ is the truncated $\ell_1$ function in \eqref{eq:tlp}. Each coefficient with absolute value at least $\tau_n$ contributes $n\lambda_n$ to the penalty. The screening step solves
\begin{equation}\label{eq:screening}
  \max_{\boldsymbol\theta}\ \ell_n^{\mathrm{pen}}(\boldsymbol\theta)
  \quad\text{subject to}\quad
  \boldsymbol\theta\in\Theta.
\end{equation}
We discuss the asymptotic conditions on the tuning parameters in Section~\ref{sec:iden} and describe their practical selection in the supplementary material.

{Problem \eqref{eq:screening} is solved by the difference-of-convex iterations of Section~\ref{subsec:pairwise} with the acyclicity constraint on the support of $B$ replaced by $h(B)=0$ in \eqref{eq:acyclicity_new}. Let $A_t=\{(j,k):|\beta^{(t)}_{jk}|\le\tau_n\}$ collect the coefficients below the truncation level at the current iterate. Each iteration maximizes $\ell_n^{\mathrm{pair}}(\boldsymbol\theta)-(n\lambda_n/\tau_n)\sum_{(j,k)\in A_t}|\beta_{jk}|$ subject to $h(B)=0$, warm-started at the current iterate, and the loop stops when $A_t$ stops updating. Each of these constrained problems is solved by} the augmented Lagrangian method \citep{Hestenes1969,nocedal2006numerical}. {It numerically optimizes a sequence of augmented objectives with increasing penalties for violations of $h(B)=0$. For each weighted $\ell_1$ problem, every limit point of the augmented-Lagrangian outer iterates is acyclic under the numerical conditions stated in the supplementary material.}
{We use $\widetilde{\boldsymbol\theta}=(\widetilde{\boldsymbol\alpha},\widetilde B,\widetilde\Omega)\in\Theta$ to denote the computed solution of the nonconvex problem \eqref{eq:screening} and define the screened DAG by}
\begin{equation}\label{eq:G0}
  \widehat G^{(0)}=\left(V,\left\{(j,k):j\ne k,\ |\widetilde\beta_{jk}|\ge\tau_n\right\}\right).
\end{equation}

We call a DAG \emph{compatible with $\Phi^*$} if it induces $\Phi^*$, and a CPDAG compatible if the DAGs in its MEC are. 
Section~\ref{sec:iden} establishes that the screened DAG is compatible with $\Phi^*$ with probability tending to one. However, unless the global solution to the highly nonconvex problem \eqref{eq:screening} is found almost exactly, this sparse DAG tends to contain spurious edges, a phenomenon explained in the theoretical analysis in 
Section~\ref{sec:iden} and further confirmed by simulation results in Section~\ref{sec:simulation}. These spurious edges will be removed by the pruning step introduced below.

\subsection{Step 2: Iterative Pruning}\label{subsec:pruning}

The iterative pruning step searches in the space of CPDAGs guided by the CLBIC. {As noted in Section~\ref{sec:model_and_problem_settings}, all DAGs in one MEC induce the same set of correlation matrices, so the maximum of $\ell_n^{\mathrm{pair}}$ over $\Theta_G$ is the same for every DAG $G$ in the MEC. For a CPDAG $\mathcal C$ we fix a DAG $G_{\mathcal C}$ in its MEC and write $\operatorname{CLBIC}_n(\mathcal C)=\operatorname{CLBIC}_n(G_{\mathcal C})$.}  For a DAG $G$, we write $\mathcal C(G)$ for its CPDAG.

{In each iteration, we consider removing one edge from a DAG in the current MEC by examining the following candidate CPDAGs:}
\begin{equation}\label{eq:pruning_neighborhood}
  \mathcal N^-(\mathcal C)
  =\bigl\{\mathcal C(G\setminus\{e\}):\ G\text{ is a DAG in the MEC of }\mathcal C\text{ and }e\text{ is an edge of }G\bigr\},
\end{equation}
where $G\setminus\{e\}$ denotes a DAG with edge $e$ removed from $G$.
This set is generated directly from $\mathcal C$ by the edge-removal operator of greedy equivalence search \citep{chickering2002equivalence,chickering2002optimal} without enumerating the DAGs in the MEC. {The pruning step is thus the backward phase of greedy equivalence search \citep{chickering2002optimal,DongEtAl2026LGES}, with the CLBIC as the score and the screened DAG $\widehat G^{(0)}$ as the starting point.}

The search starts from $\mathcal C=\mathcal C(\widehat G^{(0)})$. Each iteration computes $\operatorname{CLBIC}_n(\mathcal C')$ for every $\mathcal C'\in\mathcal N^-(\mathcal C)$ and moves to the minimizer if its CLBIC is strictly smaller than $\operatorname{CLBIC}_n(\mathcal C)$. The search stops when no candidate improves on the current CPDAG. The output is the final CPDAG $\widehat{\mathcal C}_n$ together with the parameter estimates under a DAG in its MEC. {Deleting a redundant edge changes the fit term of the CLBIC by $O_{\mathbb P}(1)$ and lowers the penalty by a multiple of $\log n$. Deleting a true edge raises the fit term by an amount of order $n$. Section~\ref{sec:iden} shows that the search therefore keeps the CPDAG compatible with $\Phi^*$ at every iteration and stops at $\mathcal C(G^*)$ with probability tending to one when it starts from a compatible $\mathcal C(\widehat G^{(0)})$. Algorithm~\ref{alg:dagclip} summarizes the two steps of DAG-CLIP.}

We employ two computational techniques to reduce the cost of the pruning step when the latent dimension $m$ is large. First, a Wald screening of the fitted coefficients ranks candidate removals. {The search evaluates candidates in this order, so the removals most likely to lower the CLBIC are evaluated first.} Second, $\widehat d_{\mathrm{eff}}$ is computed only when it can affect the comparison. The CLBIC in \eqref{eq:clbic_def} is the sum of the fit term $-2\ell_n^{\mathrm{pair}}(\widehat{\boldsymbol\theta}_G)$ and a nonnegative penalty, so the fit term alone is a lower bound on the CLBIC. If the fit term of a candidate already exceeds the smallest CLBIC found so far, the candidate is discarded without computing $\widehat d_{\mathrm{eff}}$. Neither technique changes the output of the pruning step since discarded candidates cannot attain the smallest CLBIC.

\begin{algorithm}[!t]
  \caption{DAG-CLIP}\label{alg:dagclip}
  \KwIn{Observations $\mathbf Y_1,\ldots,\mathbf Y_n$ and the assignment $j\mapsto k_j$ of manifest variables to latent variables.}
  \tcp{Step 1: penalized screening}
  {Solve \eqref{eq:screening} by difference-of-convex iterations, each by the augmented Lagrangian method with the constraint $h(B)=0$ in \eqref{eq:acyclicity_new}, and let $\widetilde{\boldsymbol\theta}$ be the solution}\;
  {Let $\widehat G^{(0)}$ be the DAG with edge set $\{(j,k):|\widetilde\beta_{jk}|\ge\tau_n\}$ as in \eqref{eq:G0}}\;
  Set $\mathcal C\leftarrow\mathcal C(\widehat G^{(0)})$ and compute $\operatorname{CLBIC}_n(\mathcal C)$\;
  \tcp{Step 2: iterative pruning}
  \Repeat{no candidate improves on $\operatorname{CLBIC}_n(\mathcal C)$}{
    Compute $\operatorname{CLBIC}_n(\mathcal C')$ for every $\mathcal C'\in\mathcal N^-(\mathcal C)$ in \eqref{eq:pruning_neighborhood}\;
    If $\min_{\mathcal C'\in\mathcal N^-(\mathcal C)}\operatorname{CLBIC}_n(\mathcal C')<\operatorname{CLBIC}_n(\mathcal C)$ then set $\mathcal C\leftarrow\argmin_{\mathcal C'\in\mathcal N^-(\mathcal C)}\operatorname{CLBIC}_n(\mathcal C')$\;
  }
  \KwOut{The CPDAG estimate $\widehat{\mathcal C}_n=\mathcal C$ and the parameter estimates under a DAG in its MEC.}
\end{algorithm}

\section{Large-sample Theory}\label{sec:iden}

This section establishes DAG-CLIP consistency for fixed $m$ and $p$ as $n\to\infty$. Let $\mathcal C^*=\mathcal C(G^*)$ denote the CPDAG of the true DAG.

\begin{assumption}[Faithfulness]\label{ass:faithfulness}
  The distribution $\mathcal N(\mathbf 0,\Phi^*)$ is faithful to $G^*$. That is, every conditional independence relationship among the components of $\boldsymbol\eta$ that holds under $\mathcal N(\mathbf 0,\Phi^*)$ is encoded by $G^*$.
\end{assumption}

{Under Assumption~\ref{ass:faithfulness}, every DAG outside the MEC of $G^*$ that induces $\Phi^*$ has strictly more edges than $G^*$ (see, Section~\ref{sec:model_and_problem_settings}).} Let $\mathfrak C_m$ be the finite set of CPDAGs on $m$ nodes. Under Assumption~\ref{ass:pairwise_id}, a CPDAG $\mathcal C\in\mathfrak C_m$ is compatible with $\Phi^*$ if and only if some $\boldsymbol\theta\in\Theta_{G_{\mathcal C}}$ reproduces all true bivariate margins, {where $G_{\mathcal C}$ is a fixed DAG in the class $\mathcal C$}. Let $\mathfrak C_0$ denote the set of compatible CPDAGs, which contains $\mathcal C^*$. Define
\[
  M_{\mathcal C}
  =\sup_{\boldsymbol\theta\in\Theta_{G_{\mathcal C}}}
  \mathbb E_*\{cl_i(\boldsymbol\theta)\},
\]
where $\mathbb E_*$ denotes expectation under the true distribution. Then $M_{\mathcal C}\le M_{\mathcal C^*}$ for every $\mathcal C\in\mathfrak C_m$, with equality if and only if $\mathcal C\in\mathfrak C_0$ {(see the supplementary material for a proof)}. For $\mathcal C\in\mathfrak C_m$ let $\boldsymbol\theta_{\mathcal C}^{\circ}$ be a maximizer of $\mathbb E_*\{cl_i(\boldsymbol\theta)\}$ over $\Theta_{G_{\mathcal C}}$ and set
\[
  H_{\mathcal C}=H(\boldsymbol\theta_{\mathcal C}^{\circ}),
  \qquad
  J_{\mathcal C}=J(\boldsymbol\theta_{\mathcal C}^{\circ}),
  \qquad
  d_{\mathrm{eff}}(\mathcal C)
  =\tr(J_{\mathcal C}H_{\mathcal C}^{-1}),
\]
the population counterparts of $\widehat H_G$, $\widehat J_G$ and $\widehat d_{\mathrm{eff}}(G)$ in \eqref{eq:clbic_def} for $G=G_{\mathcal C}$.

\begin{assumption}[Regularity]\label{ass:regularity}
  \begin{enumerate}
    \item[(a)]
          {For every DAG $G$ on $\{1,\ldots,m\}$, the parameter space $\Theta_G$ is compact, with $(\boldsymbol\alpha,\Phi)$ restricted to a common compact subset of $\mathcal A\times\mathcal S_m$ containing $(\boldsymbol\alpha^*,\Phi^*)$.} For every CPDAG {$\mathcal C\in\mathfrak C_m$}, the maximizer $\boldsymbol\theta_{\mathcal C}^{\circ}$ is unique and interior with $H_{\mathcal C}$ and $J_{\mathcal C}$ positive definite, and the standard regularity conditions for maximum composite likelihood estimation stated in the supplementary material hold for each candidate and jointly for nested pairs of candidates.

    \item[(b)]
          {The rates $\lambda_n\to0$, $n\lambda_n\to\infty$ and $\tau_n\to0$ hold, and the solution $\widetilde{\boldsymbol\theta}\in\Theta$ of the screening step satisfies}
          \[
            \ell_n^{\mathrm{pen}}(\widetilde{\boldsymbol\theta})
            \ge
            \sup_{\boldsymbol\theta\in\Theta}
            \ell_n^{\mathrm{pen}}(\boldsymbol\theta)
            -o_{\mathbb P}(n),
          \]
          and the numerical error in the computed CLBIC values is $o_{\mathbb P}(\log n)$ uniformly over the candidates.

  \end{enumerate}
\end{assumption}

{Part (a) ensures that $\widehat{\boldsymbol\theta}_{G_{\mathcal C}}$ is consistent for $\boldsymbol\theta_{\mathcal C}^{\circ}$ and that $\widehat d_{\mathrm{eff}}(\mathcal C)$ converges in probability to $d_{\mathrm{eff}}(\mathcal C)$ for every candidate. The positive definiteness of $J_{\mathcal C}$ is the standard nonsingularity condition on the variability matrix in composite likelihood inference \citep{VarinReidFirth2011}. As shown in the supplementary material, it implies that removing a redundant edge strictly reduces the effective number of parameters. The pruning step therefore does not stop at an over-specified compatible CPDAG.}
{Part (b) specifies conditions on the tuning parameters and the accuracy of the numerical solution. We take $\lambda_n=\log n/(2n)$ to give coefficients with $|\beta_{jk}|\ge\tau_n$ the usual BIC penalty of $(\log n)/2$ on the log-likelihood scale \citep{schwarz1978estimating}. This choice satisfies the required conditions on $\lambda_n$. We select $\tau_n$ by comparing the CLBIC of screened CPDAGs over a small grid of candidate truncation levels as described in the supplementary material. The screening accuracy condition concerns the numerical solution of the nonconvex screening problem. Since $0\le\ell_n^{\mathrm{pair}}-\ell_n^{\mathrm{pen}}\le n\lambda_n m(m-1)=o(n)$ on $\Theta$, this condition is equivalent to $\ell_n^{\mathrm{pair}}(\widetilde{\boldsymbol\theta})\ge\sup_{\Theta}\ell_n^{\mathrm{pair}}-o_{\mathbb P}(n)$.}

\begin{theorem}[Consistency of DAG-CLIP]\label{thm:consistency}
  Suppose Assumptions~\ref{ass:pairwise_id}, \ref{ass:faithfulness}, and
  \ref{ass:regularity} hold. Then the output of
  Algorithm~\ref{alg:dagclip} satisfies
  \begin{equation}\label{eq:consistency}
    \mathbb P(\widehat{\mathcal C}_n=\mathcal C^*)
    \longrightarrow 1
    \qquad\text{as }n\to\infty.
  \end{equation}
\end{theorem}

The same argument applies to the full-likelihood with the BIC under Assumption~\ref{ass:measurement_id}, faithfulness and the corresponding likelihood regularity conditions. This covers the linear factor model of Example~\ref{ex:linear_factor} and the directly observed nodes as in Example~\ref{ex:observed}.
Theorem~\ref{thm:consistency} follows from the two propositions below, which give the rationale of the two steps. Proofs are given in the supplementary material.

\begin{proposition}[Screening step]\label{prop:screening}
  Suppose Assumptions~\ref{ass:pairwise_id} and \ref{ass:regularity}(a) hold.
  \begin{enumerate}
    \item[(i)] {Under Assumption~\ref{ass:regularity}(b),}
          \begin{equation}\label{eq:screening_consistency}
            \mathbb P\{\mathcal C(\widehat G^{(0)})\in\mathfrak C_0\}\longrightarrow1.
          \end{equation}
    \item[(ii)] {If Assumption~\ref{ass:faithfulness} {and the tuning-parameter conditions in Assumption~\ref{ass:regularity}(b) hold} and $\ell_n^{\mathrm{pen}}(\widetilde{\boldsymbol\theta})\ge\sup_{\boldsymbol\theta\in\Theta}\ell_n^{\mathrm{pen}}(\boldsymbol\theta)-o_{\mathbb P}(n\lambda_n)$, then $$\mathbb P\{\mathcal C(\widehat G^{(0)})=\mathcal C^*\}\longrightarrow1.$$}
  \end{enumerate}
\end{proposition}

{Part (i) holds because, under Assumption~\ref{ass:regularity}(b), $\ell_n^{\mathrm{pair}}(\widetilde{\boldsymbol\theta})$ is within $o_{\mathbb P}(n)$ of its supremum, so $(\widetilde{\boldsymbol\alpha},\widetilde\Phi)$ is consistent and $\widehat G^{(0)}$ is compatible. Part (ii) requires a near-global solution of the nonconvex screening problem, which is hard to attain. Part (i) requires only that $\widetilde{\boldsymbol\theta}$ be within $o_{\mathbb P}(n)$ of the supremum, a much weaker requirement. The simulations in Section~\ref{sec:simulation} reflect this difference. The screening step alone retains redundant edges and the pruning step removes them.}

\begin{proposition}[Pruning step]\label{prop:pruning}
  Suppose Assumptions~\ref{ass:pairwise_id}, \ref{ass:faithfulness},
  and part (a) of Assumption~\ref{ass:regularity} hold. Let
  $\mathcal C\in\mathfrak C_0$.
  \begin{enumerate}
    \item[(i)] If $\mathcal C\ne\mathcal C^*$ then $\mathcal N^-(\mathcal C)\cap\mathfrak C_0\ne\varnothing$.
    \item[(ii)] If $\mathcal C'\in\mathcal N^-(\mathcal C)$ is compatible, then {$d_{\mathrm{eff}}(\mathcal C')<d_{\mathrm{eff}}(\mathcal C)$ and}
          \[
            \operatorname{CLBIC}_n(\mathcal C')-\operatorname{CLBIC}_n(\mathcal C)
            =-\{d_{\mathrm{eff}}(\mathcal C)-d_{\mathrm{eff}}(\mathcal C')\}\log n+o_{\mathbb P}(\log n).
          \]
    \item[(iii)] If $\mathcal C'\in\mathcal N^-(\mathcal C)$ is not compatible, then {$M_{\mathcal C'}<M_{\mathcal C^*}$ and}
          \[
            \operatorname{CLBIC}_n(\mathcal C')-\operatorname{CLBIC}_n(\mathcal C)
            =2n(M_{\mathcal C^*}-M_{\mathcal C'})+o_{\mathbb P}(n).
          \]
  \end{enumerate}
\end{proposition}

Proposition~\ref{prop:screening} gives a compatible starting CPDAG with
probability tending to one. {Proposition~\ref{prop:pruning}(i) gives every compatible CPDAG other than $\mathcal C^*$ a compatible neighbor \citep[Theorem 4,][]{chickering2002optimal}. A lemma in the supplementary material shows that this neighbor has a smaller effective number of parameters.} Parts (ii)
and (iii) show that a redundant deletion is favored on the $\log n$ scale,
whereas an incompatible deletion is rejected on the $n$ scale. Since
$\mathfrak C_m$ is finite and $\mathcal C^*$ is the unique edge-minimal
compatible class, the pruning step reaches and stops at $\mathcal C^*$ with
probability tending to one.

\section{Simulation Studies}\label{sec:simulation}
This section examines the finite-sample performance of the proposed DAG-CLIP. {The study evaluates latent CPDAG recovery from correlated ordinal manifest variables across three graph families, two 
latent-dimension sizes, and two sample sizes. We compare DAG-CLIP with its screening step alone (DAG-CLIP$^{(0)}$), without pruning and three two-step baselines (screening and pruning).}

\subsection{Design and Evaluation}\label{subsec:sim_design}

Each simulation draws the true DAG from one of the three graph families mentioned below with $m$ latent variables. (1)~The \textbf{random} family places $m$ edges on uniformly sampled node pairs oriented by a random topological order. (2)~The \textbf{star} family directs $\lceil(m-1)/2\rceil$ edges into a randomly chosen hub. (3)~The \textbf{tree} family arranges the $m$ nodes in a balanced out-tree with every edge directed away from the root. These three families cover sparse random structures, a concentrated hub, and a deep hierarchy, respectively, and they stress adjacency recovery and orientation to different degrees. The nonzero coefficients of $B$ are drawn independently and uniformly from $[-2.0,-0.5]\cup[0.5,2.0]$. The latent variables are then rescaled to unit variance as in Section~\ref{sec:model_and_problem_settings}, which does not change the graph or its CPDAG.

Each latent variable is measured by three ordinal manifest variables, each with five categories, following the graded response model of Example~\ref{ex:grm}, with item-specific loadings and thresholds that yield equal category probabilities. The simulation design fully crosses $m\in\{20,40\}$, (hence $p\in\{60,120\}$), with $n\in\{2000,4000\}$, resulting in four design cells with  $100$ replications each. We apply all five methods to the same generated data within each replication. Within each replication, the screening step is evaluated across truncation levels $\tau\in\{0.02,0.05,0.10\}$, using six starting values - five random and one from a Gaussian fit to the estimated latent correlation matrix. We use the screened DAG whose CPDAG has the smallest CLBIC for the pruning step.

Specifically, we compare five methods. (1)~\textbf{DAG-CLIP} is the proposed method with the settings described above. (2)~\textbf{DAG-CLIP$^{(0)}$} is the screening step of DAG-CLIP alone, included to isolate the contribution of the pruning step. The remaining three methods are two-step procedures: they first fit the graded response model by pairwise maximum likelihood, under the same assignment of manifest to latent variables and with the latent correlation matrix left unrestricted, and then apply a structure learning method for the observed variables. (3)~\textbf{Factor-score GES} (FS-GES) obtains regression factor scores and replaces each underlying response by its conditional mean given the observed category under the fitted parameters, and then applies greedy equivalence search \citep{chickering2002optimal} with the Gaussian BIC to the resulting $n\times m$ matrix of scores. (4)~\textbf{Latent-correlation GES} (LC-GES) applies the same search procedure directly to the estimated latent correlation matrix, treating it as the sample covariance matrix of a Gaussian sample of size $n$. (5)~\textbf{Factor-score PC} (FS-PC) applies the PC algorithm \citep{SpirtesGlymourScheines2000} to the factor scores, using Fisher-$z$ tests at level $0.01$. Each test checks whether two factor scores are independent after adjusting for up to four other factor scores. Each baseline treats an estimated summary of the measurement model as observed data. Further details of the simulation design and the baseline methods are given in the supplementary material.

We evaluate the five methods above, including the proposed method, using five common accuracy measures from causal structure learning. All five measures compare the estimated and true CPDAGs. For a CPDAG $\mathcal C$ let $\operatorname{skel}(\mathcal C)$ denote its set of adjacencies, the unordered pairs of nodes joined by an edge, and $\operatorname{arr}(\mathcal C)$ its set of ordered pairs $(j,k)$ for which $j\to k$ is a directed edge in $\mathcal C$.

For an estimated CPDAG $\widehat{\mathcal C}$ the CPDAG structural Hamming distance is
\[
  \operatorname{SHD}(\widehat{\mathcal C},\mathcal C^*)
  =\sum_{1\le j<k\le m}1\{e_{jk}(\widehat{\mathcal C})\ne e_{jk}(\mathcal C^*)\},
\]
where $e_{jk}(\mathcal C)$ records whether nodes $j$ and $k$ are nonadjacent, joined by $j\to k$, joined by $k\to j$ or joined by an undirected edge in $\mathcal C$. 

The skeleton F1 and the compelled-arrow F1 are
\[
  F_1^{\mathrm{skel}}
  =\frac{2\,|\operatorname{skel}(\widehat{\mathcal C})\cap\operatorname{skel}(\mathcal C^*)|}
  {|\operatorname{skel}(\widehat{\mathcal C})|+|\operatorname{skel}(\mathcal C^*)|},
  \qquad
  F_1^{\mathrm{arrow}}
  =\frac{2\,|\operatorname{arr}(\widehat{\mathcal C})\cap\operatorname{arr}(\mathcal C^*)|}
  {|\operatorname{arr}(\widehat{\mathcal C})|+|\operatorname{arr}(\mathcal C^*)|},
\]
which measure adjacency and orientation accuracy, respectively, and exact recovery is the indicator $1\{\widehat{\mathcal C}=\mathcal C^*\}$ denoted as MEC in the result Tables. We also report Extra, the number of estimated redundant adjacencies that are not in the true graph. {We average all five quantities over the $100$ independent replications of each setting. Each method is allowed 24 hours of computation per replication. FS-GES and LC-GES exceed this limit in some settings at $m=40$, and their averages use only the replications that finished. Table~\ref{tab:ordinal_recovery_m40} marks these settings and gives the numbers of finished replications.}

\begin{table}[t]
  \centering
  \footnotesize
  \caption{Ordinal latent-MEC recovery at $m=20$ {over $100$ replications for each setting.} Entries are the mean CPDAG structural Hamming distance (SHD),
    skeleton F1 (Skel.), compelled-arrow F1 (Arrow), exact-MEC recovery
    proportion (MEC) and the mean number of estimated adjacencies that are
    not in the true graph (Extra), with standard errors in parentheses. The
    best value in each column of a panel is in bold. Arrow F1 is omitted for the tree family that has no compelled arrows.}
  \label{tab:ordinal_recovery_m20}
  \setlength{\tabcolsep}{2pt}
  \begingroup\scriptsize
  \resizebox{\linewidth}{!}{\begin{tabular}{l rrrrr rrrrr}
    \toprule
             & \multicolumn{5}{c}{$n{=}2000$}   & \multicolumn{5}{c}{$n{=}4000$}                                                                                                                                                                                                                                  \\
    \cmidrule(lr){2-6}\cmidrule(lr){7-11}
    Method & \multicolumn{1}{c}{SHD} & \multicolumn{1}{c}{Skel.} & \multicolumn{1}{c}{Arrow} & \multicolumn{1}{c}{MEC} & \multicolumn{1}{c}{Extra} & \multicolumn{1}{c}{SHD} & \multicolumn{1}{c}{Skel.} & \multicolumn{1}{c}{Arrow} & \multicolumn{1}{c}{MEC} & \multicolumn{1}{c}{Extra} \\
    \midrule
    \multicolumn{11}{l}{\emph{Panel A. Random}}                                                                                                                                                                                                                                                                    \\
    DAG-CLIP & \textbf{5.3}\,{\scriptsize(0.5)} & \textbf{0.89}\,{\scriptsize(0.01)} & \textbf{0.82}\,{\scriptsize(0.02)} & \textbf{0.20}\,{\scriptsize(0.04)} & \textbf{2.8}\,{\scriptsize(0.3)} & \textbf{3.5}\,{\scriptsize(0.4)} & \textbf{0.93}\,{\scriptsize(0.01)} & \textbf{0.89}\,{\scriptsize(0.01)} & \textbf{0.41}\,{\scriptsize(0.05)} & \textbf{1.9}\,{\scriptsize(0.3)} \\
    DAG-CLIP$^{(0)}$ & 10.5\,{\scriptsize(0.7)} & 0.81\,{\scriptsize(0.01)} & 0.72\,{\scriptsize(0.02)} & 0.06\,{\scriptsize(0.02)} & 7.3\,{\scriptsize(0.5)} & 6.9\,{\scriptsize(0.7)} & 0.87\,{\scriptsize(0.01)} & 0.83\,{\scriptsize(0.02)} & 0.24\,{\scriptsize(0.04)} & 5.0\,{\scriptsize(0.5)} \\
    FS-PC & 25.2\,{\scriptsize(0.9)} & 0.67\,{\scriptsize(0.01)} & 0.36\,{\scriptsize(0.02)} & 0.00\,{\scriptsize(0.00)} & 14.5\,{\scriptsize(0.6)} & 26.0\,{\scriptsize(0.9)} & 0.66\,{\scriptsize(0.01)} & 0.39\,{\scriptsize(0.02)} & 0.00\,{\scriptsize(0.00)} & 16.1\,{\scriptsize(0.6)} \\
    FS-GES & 89.4\,{\scriptsize(3.1)} & 0.34\,{\scriptsize(0.01)} & 0.22\,{\scriptsize(0.01)} & 0.00\,{\scriptsize(0.00)} & 80.9\,{\scriptsize(2.9)} & 86.6\,{\scriptsize(2.9)} & 0.35\,{\scriptsize(0.01)} & 0.24\,{\scriptsize(0.01)} & 0.00\,{\scriptsize(0.00)} & 79.5\,{\scriptsize(2.8)} \\
    LC-GES & 78.0\,{\scriptsize(3.2)} & 0.38\,{\scriptsize(0.01)} & 0.25\,{\scriptsize(0.01)} & 0.00\,{\scriptsize(0.00)} & 70.0\,{\scriptsize(3.0)} & 76.7\,{\scriptsize(3.2)} & 0.39\,{\scriptsize(0.01)} & 0.27\,{\scriptsize(0.01)} & 0.00\,{\scriptsize(0.00)} & 69.7\,{\scriptsize(3.1)} \\
    \midrule
    \multicolumn{11}{l}{\emph{Panel B. Star}}                                                                                                                                                                                                                                                                      \\
    DAG-CLIP & \textbf{0.7}\,{\scriptsize(0.1)} & \textbf{0.97}\,{\scriptsize(0.00)} & \textbf{0.98}\,{\scriptsize(0.00)} & \textbf{0.55}\,{\scriptsize(0.05)} & \textbf{0.5}\,{\scriptsize(0.1)} & \textbf{0.3}\,{\scriptsize(0.1)} & \textbf{0.98}\,{\scriptsize(0.00)} & \textbf{1.00}\,{\scriptsize(0.00)} & \textbf{0.71}\,{\scriptsize(0.05)} & \textbf{0.3}\,{\scriptsize(0.1)} \\
    DAG-CLIP$^{(0)}$ & 4.4\,{\scriptsize(0.3)} & 0.84\,{\scriptsize(0.01)} & 0.91\,{\scriptsize(0.01)} & 0.02\,{\scriptsize(0.01)} & 4.2\,{\scriptsize(0.3)} & 3.2\,{\scriptsize(0.2)} & 0.87\,{\scriptsize(0.01)} & 0.94\,{\scriptsize(0.01)} & 0.08\,{\scriptsize(0.03)} & 3.2\,{\scriptsize(0.2)} \\
    FS-PC & 10.7\,{\scriptsize(0.3)} & 0.69\,{\scriptsize(0.01)} & 0.63\,{\scriptsize(0.01)} & 0.00\,{\scriptsize(0.00)} & 9.1\,{\scriptsize(0.3)} & 10.8\,{\scriptsize(0.4)} & 0.69\,{\scriptsize(0.01)} & 0.65\,{\scriptsize(0.01)} & 0.00\,{\scriptsize(0.00)} & 9.5\,{\scriptsize(0.3)} \\
    FS-GES & 14.2\,{\scriptsize(0.4)} & 0.59\,{\scriptsize(0.01)} & 0.66\,{\scriptsize(0.01)} & 0.00\,{\scriptsize(0.00)} & 14.2\,{\scriptsize(0.4)} & 12.2\,{\scriptsize(0.3)} & 0.63\,{\scriptsize(0.01)} & 0.70\,{\scriptsize(0.01)} & 0.00\,{\scriptsize(0.00)} & 12.2\,{\scriptsize(0.3)} \\
    LC-GES & 8.6\,{\scriptsize(0.3)} & 0.71\,{\scriptsize(0.01)} & 0.78\,{\scriptsize(0.01)} & 0.00\,{\scriptsize(0.00)} & 8.6\,{\scriptsize(0.3)} & 6.6\,{\scriptsize(0.2)} & 0.76\,{\scriptsize(0.01)} & 0.84\,{\scriptsize(0.01)} & 0.00\,{\scriptsize(0.00)} & 6.6\,{\scriptsize(0.2)} \\
    \midrule
    \multicolumn{11}{l}{\emph{Panel C. Tree}}                                                                                                                                                                                                                                                                      \\
    DAG-CLIP & \textbf{3.7}\,{\scriptsize(0.5)} & \textbf{0.95}\,{\scriptsize(0.01)} & \multicolumn{1}{c}{--} & \textbf{0.47}\,{\scriptsize(0.05)} & \textbf{1.7}\,{\scriptsize(0.2)} & \textbf{3.5}\,{\scriptsize(0.6)} & \textbf{0.94}\,{\scriptsize(0.01)} & \multicolumn{1}{c}{--} & \textbf{0.60}\,{\scriptsize(0.05)} & \textbf{1.7}\,{\scriptsize(0.3)} \\
    DAG-CLIP$^{(0)}$ & 16.2\,{\scriptsize(0.9)} & 0.79\,{\scriptsize(0.01)} & \multicolumn{1}{c}{--} & 0.01\,{\scriptsize(0.01)} & 10.1\,{\scriptsize(0.6)} & 13.7\,{\scriptsize(1.1)} & 0.81\,{\scriptsize(0.01)} & \multicolumn{1}{c}{--} & 0.11\,{\scriptsize(0.03)} & 8.9\,{\scriptsize(0.8)} \\
    FS-PC & 38.4\,{\scriptsize(0.6)} & 0.59\,{\scriptsize(0.01)} & \multicolumn{1}{c}{--} & 0.00\,{\scriptsize(0.00)} & 21.6\,{\scriptsize(0.5)} & 41.5\,{\scriptsize(0.6)} & 0.58\,{\scriptsize(0.01)} & \multicolumn{1}{c}{--} & 0.00\,{\scriptsize(0.00)} & 24.0\,{\scriptsize(0.6)} \\
    FS-GES & 131.0\,{\scriptsize(2.7)} & 0.26\,{\scriptsize(0.01)} & \multicolumn{1}{c}{--} & 0.00\,{\scriptsize(0.00)} & 114.4\,{\scriptsize(2.7)} & 129.8\,{\scriptsize(2.8)} & 0.26\,{\scriptsize(0.01)} & \multicolumn{1}{c}{--} & 0.00\,{\scriptsize(0.00)} & 113.1\,{\scriptsize(2.8)} \\
    LC-GES & 115.0\,{\scriptsize(2.6)} & 0.29\,{\scriptsize(0.01)} & \multicolumn{1}{c}{--} & 0.00\,{\scriptsize(0.00)} & 98.9\,{\scriptsize(2.5)} & 116.1\,{\scriptsize(2.8)} & 0.29\,{\scriptsize(0.01)} & \multicolumn{1}{c}{--} & 0.00\,{\scriptsize(0.00)} & 99.9\,{\scriptsize(2.7)} \\
    \bottomrule
  \end{tabular}}
  \endgroup
\end{table}

\begin{table}[t]
  \centering
  \footnotesize
  \caption{Ordinal latent-MEC recovery at $m=40$. {Columns and bold entries are as in Table~\ref{tab:ordinal_recovery_m20}. $^\dagger$ marks settings in which FS-GES or LC-GES did not finish all $100$ replications within the 24-hour limit. These entries are averages over the finished replications, with counts for $n=2000$ and $n=4000$ in parentheses.}}
  \label{tab:ordinal_recovery_m40}
  \setlength{\tabcolsep}{2pt}
  \begingroup\scriptsize
  \resizebox{\linewidth}{!}{\begin{tabular}{l rrrrr rrrrr}
    \toprule
                                          & \multicolumn{5}{c}{$n{=}2000$}    & \multicolumn{5}{c}{$n{=}4000$}                                                                                                                                                                                                                         \\
    \cmidrule(lr){2-6}\cmidrule(lr){7-11}
    Method & \multicolumn{1}{c}{SHD} & \multicolumn{1}{c}{Skel.} & \multicolumn{1}{c}{Arrow} & \multicolumn{1}{c}{MEC} & \multicolumn{1}{c}{Extra} & \multicolumn{1}{c}{SHD} & \multicolumn{1}{c}{Skel.} & \multicolumn{1}{c}{Arrow} & \multicolumn{1}{c}{MEC} & \multicolumn{1}{c}{Extra} \\
    \midrule
    \multicolumn{11}{l}{\emph{Panel A. Random}}                                                                                                                                                                                                                                                                                         \\
    DAG-CLIP & \textbf{14.9}\,{\scriptsize(0.8)} & \textbf{0.84}\,{\scriptsize(0.01)} & \textbf{0.76}\,{\scriptsize(0.01)} & 0.00\,{\scriptsize(0.00)} & \textbf{8.2}\,{\scriptsize(0.5)} & \textbf{11.1}\,{\scriptsize(0.9)} & \textbf{0.88}\,{\scriptsize(0.01)} & \textbf{0.82}\,{\scriptsize(0.01)} & \textbf{0.06}\,{\scriptsize(0.02)} & \textbf{6.3}\,{\scriptsize(0.5)} \\
    DAG-CLIP$^{(0)}$ & 33.7\,{\scriptsize(1.4)} & 0.71\,{\scriptsize(0.01)} & 0.62\,{\scriptsize(0.01)} & 0.00\,{\scriptsize(0.00)} & 25.3\,{\scriptsize(1.2)} & 27.1\,{\scriptsize(1.4)} & 0.76\,{\scriptsize(0.01)} & 0.69\,{\scriptsize(0.01)} & 0.00\,{\scriptsize(0.00)} & 20.6\,{\scriptsize(1.0)} \\
    FS-PC & 64.5\,{\scriptsize(1.8)} & 0.59\,{\scriptsize(0.01)} & 0.30\,{\scriptsize(0.01)} & 0.00\,{\scriptsize(0.00)} & 41.3\,{\scriptsize(1.4)} & 69.4\,{\scriptsize(1.7)} & 0.58\,{\scriptsize(0.01)} & 0.28\,{\scriptsize(0.01)} & 0.00\,{\scriptsize(0.00)} & 46.0\,{\scriptsize(1.3)} \\
    FS-GES {\scriptsize(81/75)}$^\dagger$ & 369.9\,{\scriptsize(8.1)} & 0.19\,{\scriptsize(0.00)} & 0.13\,{\scriptsize(0.00)} & 0.00\,{\scriptsize(0.00)} & 356.3\,{\scriptsize(8.0)} & 351.8\,{\scriptsize(10.0)} & 0.20\,{\scriptsize(0.01)} & 0.14\,{\scriptsize(0.01)} & 0.00\,{\scriptsize(0.00)} & 338.6\,{\scriptsize(9.9)} \\
    LC-GES {\scriptsize(94/88)}$^\dagger$ & 346.1\,{\scriptsize(8.9)} & 0.20\,{\scriptsize(0.01)} & 0.14\,{\scriptsize(0.01)} & 0.00\,{\scriptsize(0.00)} & 331.7\,{\scriptsize(8.8)} & 328.2\,{\scriptsize(9.8)} & 0.21\,{\scriptsize(0.01)} & 0.15\,{\scriptsize(0.01)} & 0.00\,{\scriptsize(0.00)} & 314.4\,{\scriptsize(9.7)} \\
    \midrule
    \multicolumn{11}{l}{\emph{Panel B. Star}}                                                                                                                                                                                                                                                                                           \\
    DAG-CLIP & \textbf{4.7}\,{\scriptsize(0.2)} & \textbf{0.89}\,{\scriptsize(0.01)} & \textbf{0.94}\,{\scriptsize(0.01)} & \textbf{0.03}\,{\scriptsize(0.02)} & \textbf{3.0}\,{\scriptsize(0.2)} & \textbf{3.0}\,{\scriptsize(0.2)} & \textbf{0.93}\,{\scriptsize(0.00)} & \textbf{0.96}\,{\scriptsize(0.00)} & \textbf{0.08}\,{\scriptsize(0.03)} & \textbf{1.8}\,{\scriptsize(0.1)} \\
    DAG-CLIP$^{(0)}$ & 21.6\,{\scriptsize(0.6)} & 0.64\,{\scriptsize(0.01)} & 0.72\,{\scriptsize(0.01)} & 0.00\,{\scriptsize(0.00)} & 19.9\,{\scriptsize(0.6)} & 20.0\,{\scriptsize(0.7)} & 0.66\,{\scriptsize(0.01)} & 0.74\,{\scriptsize(0.01)} & 0.00\,{\scriptsize(0.00)} & 18.9\,{\scriptsize(0.7)} \\
    FS-PC & 39.5\,{\scriptsize(0.6)} & 0.53\,{\scriptsize(0.00)} & 0.42\,{\scriptsize(0.01)} & 0.00\,{\scriptsize(0.00)} & 34.3\,{\scriptsize(0.5)} & 38.9\,{\scriptsize(0.7)} & 0.54\,{\scriptsize(0.00)} & 0.44\,{\scriptsize(0.01)} & 0.00\,{\scriptsize(0.00)} & 34.2\,{\scriptsize(0.5)} \\
    FS-GES & 51.8\,{\scriptsize(0.8)} & 0.44\,{\scriptsize(0.00)} & 0.46\,{\scriptsize(0.00)} & 0.00\,{\scriptsize(0.00)} & 51.8\,{\scriptsize(0.8)} & 43.3\,{\scriptsize(0.7)} & 0.48\,{\scriptsize(0.00)} & 0.52\,{\scriptsize(0.01)} & 0.00\,{\scriptsize(0.00)} & 43.3\,{\scriptsize(0.7)} \\
    LC-GES & 32.1\,{\scriptsize(0.6)} & 0.56\,{\scriptsize(0.00)} & 0.61\,{\scriptsize(0.01)} & 0.00\,{\scriptsize(0.00)} & 32.1\,{\scriptsize(0.6)} & 25.1\,{\scriptsize(0.5)} & 0.62\,{\scriptsize(0.01)} & 0.69\,{\scriptsize(0.01)} & 0.00\,{\scriptsize(0.00)} & 25.1\,{\scriptsize(0.5)} \\
    \midrule
    \multicolumn{11}{l}{\emph{Panel C. Tree}}                                                                                                                                                                                                                                                                                           \\
    DAG-CLIP & \textbf{14.6}\,{\scriptsize(1.4)} & \textbf{0.88}\,{\scriptsize(0.01)} & \multicolumn{1}{c}{--} & \textbf{0.06}\,{\scriptsize(0.02)} & \textbf{7.3}\,{\scriptsize(0.7)} & \textbf{11.7}\,{\scriptsize(1.4)} & \textbf{0.90}\,{\scriptsize(0.01)} & \multicolumn{1}{c}{--} & \textbf{0.16}\,{\scriptsize(0.04)} & \textbf{6.6}\,{\scriptsize(0.8)} \\
    DAG-CLIP$^{(0)}$ & 57.4\,{\scriptsize(2.1)} & 0.66\,{\scriptsize(0.01)} & \multicolumn{1}{c}{--} & 0.00\,{\scriptsize(0.00)} & 36.8\,{\scriptsize(1.5)} & 62.4\,{\scriptsize(2.7)} & 0.64\,{\scriptsize(0.01)} & \multicolumn{1}{c}{--} & 0.00\,{\scriptsize(0.00)} & 42.0\,{\scriptsize(2.1)} \\
    FS-PC & 77.5\,{\scriptsize(0.9)} & 0.60\,{\scriptsize(0.01)} & \multicolumn{1}{c}{--} & 0.00\,{\scriptsize(0.00)} & 43.2\,{\scriptsize(0.8)} & 90.4\,{\scriptsize(0.9)} & 0.54\,{\scriptsize(0.01)} & \multicolumn{1}{c}{--} & 0.00\,{\scriptsize(0.00)} & 54.2\,{\scriptsize(0.8)} \\
    FS-GES {\scriptsize(52/30)}$^\dagger$ & 280.9\,{\scriptsize(6.0)} & 0.24\,{\scriptsize(0.00)} & \multicolumn{1}{c}{--} & 0.00\,{\scriptsize(0.00)} & 245.4\,{\scriptsize(6.0)} & 298.7\,{\scriptsize(7.1)} & 0.23\,{\scriptsize(0.00)} & \multicolumn{1}{c}{--} & 0.00\,{\scriptsize(0.00)} & 262.7\,{\scriptsize(6.9)} \\
    LC-GES {\scriptsize(69/33)}$^\dagger$ & 276.4\,{\scriptsize(5.3)} & 0.24\,{\scriptsize(0.00)} & \multicolumn{1}{c}{--} & 0.00\,{\scriptsize(0.00)} & 240.7\,{\scriptsize(5.3)} & 282.8\,{\scriptsize(7.3)} & 0.24\,{\scriptsize(0.01)} & \multicolumn{1}{c}{--} & 0.00\,{\scriptsize(0.00)} & 247.0\,{\scriptsize(7.2)} \\
    \bottomrule                      
  \end{tabular}}
  \endgroup
\end{table}

\subsection{Results}\label{subsec:sim_categorical}

Tables~\ref{tab:ordinal_recovery_m20} and \ref{tab:ordinal_recovery_m40} report the five accuracy measures by graph family and sample size, for $m=20$ and $m=40$, respectively. 
Overall, DAG-CLIP achieves high graph recovery accuracy and outperforms FS-PC, FS-GES, and LC-GES across the simulated settings.
We further summarize four main findings. {First, for every combination of graph family, latent dimension, and sample size, DAG-CLIP achieves the lowest average SHD and highest average skeleton F1 across replications. It also achieves the highest average compelled-arrow F1 in the random and star families.} Averaged over the three families, its SHD is about one tenth of that of FS-PC, the best baseline, at $m=20$ and about one sixth at $m=40$. Furthermore, exact recovery requires every adjacency and orientation to be correct. {Both DAG-CLIP and DAG-CLIP$^{(0)}$ achieve positive exact-recovery rates, whereas the three two-step baselines never recover the true class exactly. For DAG-CLIP at $m=20$ and $n=4000$}, the proportion is $0.71$ for the star family, $0.60$ for the tree family and $0.41$ for the random family. As the latent dimension increases from 20 to 40, DAG-CLIP's performance remains very good and much better than the baselines, but it slightly worsens. 

Second, sample size affects DAG-CLIP and the three baseline methods differently. Doubling $n$ {generally improves the average performance of DAG-CLIP}, while the baseline methods do not always improve and can even get worse. In particular, FS-PC performs worse on most measures in the random and tree graph families, and {the GES baselines generally perform better on random and star graphs but worse on tree graphs}. This difference stems from two error sources. The baselines treat factor scores or an estimated correlation matrix as observed data, and the resulting bias does not vanish as $n$ grows. The error of DAG-CLIP is dominated by sampling variability and decreases with $n$ as Theorem~\ref{thm:consistency} shows.

Third, graph families differ in how well the five methods recover adjacency and orientation. In the star family, every edge is oriented by the v-structures at the hub. DAG-CLIP recovers almost all of these orientations, while the baselines recover far fewer. {In the tree family, no edge is oriented. The baselines have their largest SHD in this family except for the GES baselines at $m=40$, whose largest SHD occurs in the random family.} DAG-CLIP {has an average skeleton F1 of at least $0.88$ in every tree setting}. The two GES baselines do not complete all random and tree replications at $m=40$ {within the 24-hour limit}.

{Fourth, comparison with DAG-CLIP$^{(0)}$ in Tables~\ref{tab:ordinal_recovery_m20} and \ref{tab:ordinal_recovery_m40} shows that pruning removes most of the extra adjacencies retained by screening alone. It reduces SHD and increases skeleton F1 in every setting with higher exact-recovery rates.}

\section{Application to TIMSS 2023}\label{sec:real}
\suppressfloats[t]

\subsection{Data and Graph Nodes}\label{subsec:timss_preparation}

The Trends in International Mathematics and Science Study (TIMSS) is an international assessment administered by the International Association for the Evaluation of Educational Achievement (IEA) that aims to measure how well students in Year 4 and Year 8 have mastered the knowledge they are expected to learn as part of the mathematics and science curriculum. The TIMSS questionnaire also collects information on various aspects of education, such as home resources, attitudes, and classroom experiences \citep{vonDavierFishbeinKennedy2024}. Overall, the availability of constructs and background variables in TIMSS provides many opportunities for education research. 

We focus on fourteen latent constructs measured by the TIMSS student questionnaire and two observed achievement scores. Broadly, these variables cover mathematics- and science-specific achievement; attitudes and motivation; classroom and instructional environment; school climate and social experiences; home resources and student background context; and digital literacy. Researchers are often interested not only in how student-specific characteristics and attitudes affect achievement but also in how constructs such as classroom and teacher-specific variables correlate with each other.

Contrary to testing specific education hypotheses, our analysis, which is enabled by our proposed framework, is exploratory and focuses on two research questions. The first asks which relationships among the sixteen graph nodes {are represented by adjacencies in the fitted graph} (in other words, which edges exist). The second asks which of their directions {are shared by all DAGs in the selected MEC} (for which directions there is compelling evidence). These are the adjacencies and the directed edges of the CPDAG introduced in Section~\ref{sec:model_and_problem_settings}. We answer these two questions by implementing the proposed methodology in building the CPDAG.

Because language translations, item parameters, curricula and sampling designs differ across education systems, we analyze data from one country, specifically Australia, Grade 8 public-use data of TIMSS 2023 \citep{IEATIMSS2023G8,FishbeinTanevaKowolik2025,vonDavierFishbeinKennedy2024}. The analysis consists of 7,997 students from 276 schools who responded to the context questionnaire.

\begin{table}[t]
  \centering
  \small
  \caption{{TIMSS measurement configuration. In parentheses, we give node abbreviations and the number of indicators for each latent construct. The Count column gives the number of nodes per group. Achievement scores are observed.}}
  \label{tab:timss_nodes}
  \begin{tabular}{@{}>{\raggedright\arraybackslash}p{0.18\linewidth}>{\raggedright\arraybackslash}p{0.6\linewidth}>{\centering\arraybackslash}p{0.06\linewidth}>{\centering\arraybackslash}p{0.06\linewidth}@{}}
    \toprule
    Group & Nodes                                                                                                                                                                                                            & Count & Items \\
    \midrule
    General resources and experiences
          & Home Educational Resources (HER; 4), Digital Self-Efficacy (DSE; 5), Sense of School Belonging (SSB; 5), Bullying Exposure (BULL; 5)
          & 4                                                                                                                                                                                                                & 19            \\
    Mathematics context
          & Like Learning Mathematics (LLM; 5), Instructional Clarity in Mathematics Lessons (ICM; 5), Disorderly Behavior During Mathematics Lessons (DML; 5), Confidence in Mathematics (CM; 5), Value Mathematics (VM; 5)
          & 5                                                                                                                                                                                                                & 25            \\
    Science context
          & Like Learning Science (LLS; 5), Instructional Clarity in Science Lessons (ICS; 5), Disorderly Behavior During Science Lessons (DLS; 5), Confidence in Science (CS; 5), Value Science (VS; 5)
          & 5                                                                                                                                                                                                                & 25            \\
    Achievement (observed scores)
          & Mathematics Achievement (MA), Science Achievement (SA)
          & 2                                                                                                                                                                                                                & 2             \\
    \midrule
    Total &                                                                                                                                                                                                                  & 16    & 71    \\
    \bottomrule
  \end{tabular}
\end{table}

Table~\ref{tab:timss_nodes} lists the 16 graph nodes. Fourteen latent constructs are measured by 69 questionnaire items (each construct is measured by a small block of binary or ordered categorical items). 
Specifically, four constructs describe general resources and experiences, and two parallel groups of five constructs describe the mathematics and the science learning context. The remaining two nodes are mathematics and science achievement. For the achievement constructs, TIMSS reports achievement through five plausible values per student \citep{vonDavierFishbeinKennedy2024}. For each student, we average the five plausible values into one score, standardize it and enter it as an observed node as in Example~\ref{ex:observed}. Regarding the remaining constructs, the item blocks follow the published TIMSS scale assignments, so each item measures a single construct as assumed in Section~\ref{sec:model_and_problem_settings}. {
  To keep the number of item pairs manageable, we fit a one-factor probit model to each block and retain the five items with the largest loadings. Home Educational Resources has only four items, and we retain all of them.}
Every latent construct therefore has at least three indicators and satisfies the identifiability conditions discussed in Sections~\ref{sec:model_and_problem_settings} and \ref{subsec:pairwise}. {The supplementary material describes the item blocks and their coding.}

For model fitting and testing, we randomly split students into a design sample (20\%), a fitting sample (60\%), and a test sample (20\%). {We use the design sample to merge sparse categories, select items from each item block through the one-factor model as described above, and choose one anchor item per construct (set the loading of one item in each block of items to one to fix the scale of the latent variables). Item response categories are coded so that a higher response category indicates a higher position on the corresponding construct.
{The analysis includes three binary items, 66 ordered polytomous items with 3 to 5 response categories, and two achievement scores.}
Since all item indicators within a block of items measure the same construct and are positively correlated, all loadings are positive.} {We use the fitting sample to estimate the model and select the graph with DAG-CLIP, and the test sample only for out-of-sample evaluation.} The analysis uses complete cases, as the retained items have nonresponse rates between one and five per cent. 

{We treat the resources and experience constructs that capture students' home and personal resources as well as broader school experiences as background conditions for subject-specific learning contexts and achievement. {We therefore exclude arrows from achievement and learning-context constructs into this background group, while leaving within-group directions unrestricted.} The target is a MEC consistent with this restriction. No other directions are constrained in the search. The selected orientations are discussed in Section~\ref{subsec:timss_interpretation}.}

\begin{figure}[t]
  \centering
  \includegraphics[width=.85\linewidth]{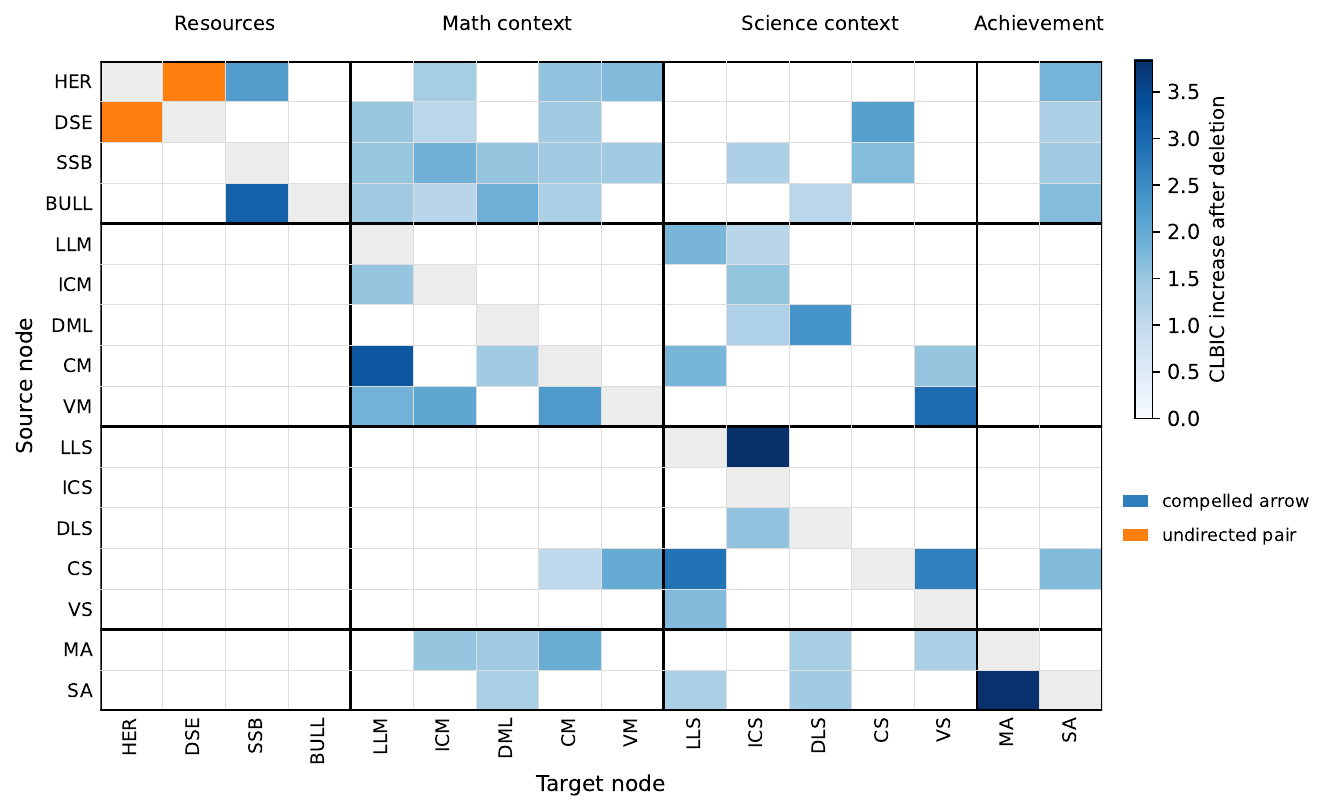}
  \caption{{Selected CPDAG. Blue cells indicate arrows from the row variable to the column variable. Darker blue indicates a larger CLBIC increase on deleting the adjacency. Orange marks the undirected HER--DSE edge.}}
  \label{fig:timss_matrix}
\end{figure}

\subsection{Selected Structure and Held-Out Prediction}\label{subsec:timss_structure}\label{subsec:timss_prediction}

We apply DAG-CLIP to the 3,110 students in the fitting sample.  Detailed estimation settings are reported in the supplementary material. The selected CPDAG has 57 of the 120 possible adjacencies. 56 are directed, and one edge between HER and DSE is undirected. Figure~\ref{fig:timss_matrix} displays the graph as an adjacency matrix. Each cell is shaded by the CLBIC increase caused by deleting the adjacency. Every retained adjacency has a positive increase, indicating retention in these model comparisons. The increases range from 0.002 to 3.83, with a median of 0.17. The two largest are 3.83 for ``Like Learning Science'' (LLS) $\to$ ``Instructional Clarity in Science Lessons'' (ICS) and 3.75 for ``Science Achievement'' (SA) $\to$ ``Mathematics Achievement'' (MA). {ICS measures students' perceptions of instructional clarity. The LLS $\to$ ICS direction, compelled within the selected MEC, suggests that greater enjoyment of science may lead students to perceive their teacher's explanations as clearer.} The edge SA $\to$ MA is discussed in Section~\ref{subsec:timss_interpretation}. The supplementary material lists the standardized coefficients and deletion increases for all retained edges, together with their CPDAG directions.

\begin{figure}[t]
  \centering
  \includegraphics[width=\linewidth]{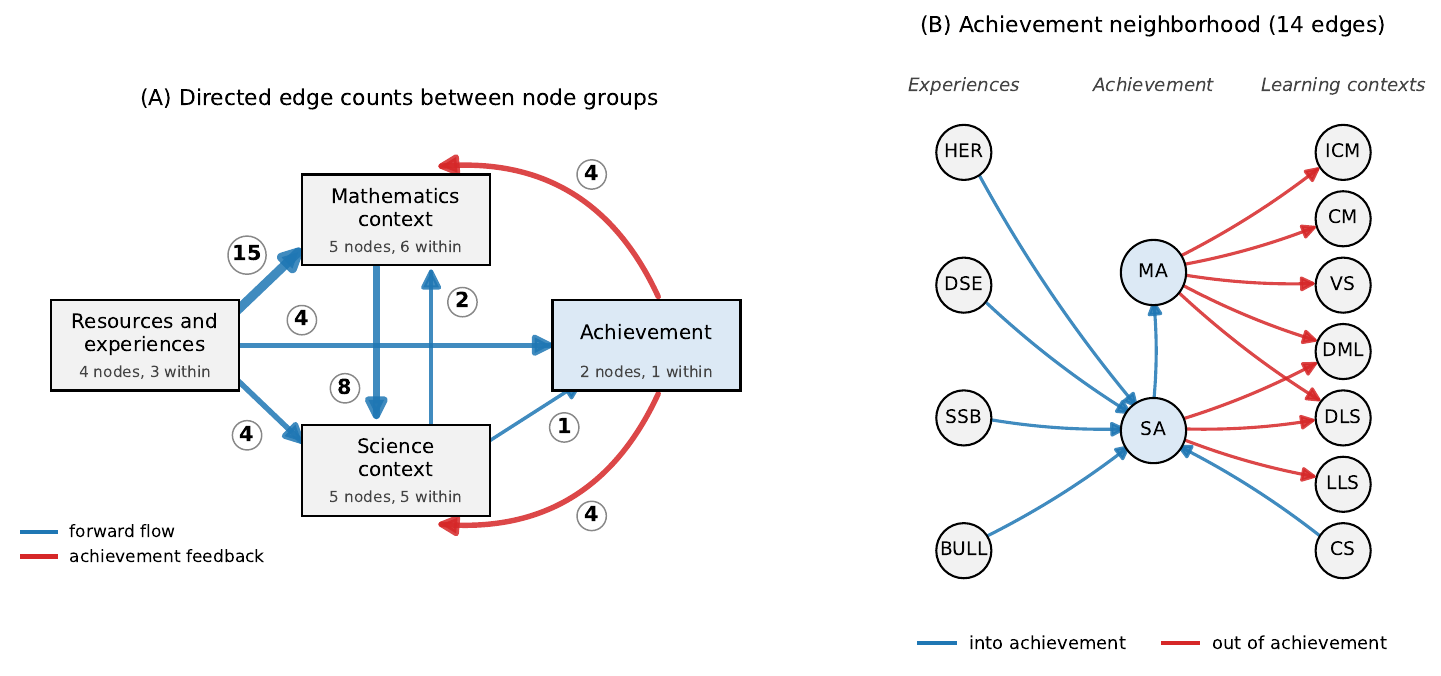}
  \caption{Summary display of the selected structure. (A) Directed edge counts between the node groups, with red arcs marking the arrows from the achievement scores {into} the learning contexts. Each box reports its node count and its number of within-group edges. (B) The 14 edges incident to the achievement scores, with blue arrows into and red arrows out of achievement.}
  \label{fig:timss_summary}
\end{figure}

Figure~\ref{fig:timss_summary} summarizes the structure at the group level and around the two achievement scores. Three features stand out. First, the four resource and experience constructs, which the restriction allows only as sources {of between-group arrows}, send 23 edges into the other groups, and nine of the twelve remaining constructs have at least one of them as a direct parent. Second, edges between the two learning contexts run mainly from mathematics to science, with eight in that direction and two in the reverse. Each mathematics construct points into its science counterpart, except that ``Confidence in Science'' (CS) points into ``Confidence in Mathematics'' (CM). Third, achievement in mathematics and science receives five edges and sends eight into the learning contexts. Specifically, ``Home Educational Resources'' (HER), ``Digital Self-Efficacy'' (DSE), ``Sense of School Belonging'' (SSB), ``Bullying Ex\-posure'' (BULL) and ``Confidence in Science'' (CS) point into ``Science Achievement'' (SA), and SA is the only parent of ``Mathematics Achievement'' (MA).

{Table~\ref{tab:timss_heldout} reports the mean and SD of the per-student pairwise negative log-likelihood (NLL) for the 1,063 test students. Both NLL and $\widehat d_{\mathrm{eff}}$ use the pairwise log-likelihood averaged over variable pairs.} {Compared with DAG-CLIP, the} empty graph, which removes all dependence among the latent variables, increases the mean loss by $21.9\times10^{-3}$ with standard error $1.4\times10^{-3}$. The saturated model with all 120 adjacencies lowers it by only $0.039\times10^{-3}$ with standard error $0.020\times10^{-3}$. The selected CPDAG thus {achieves 99.8 percent of the saturated model's reduction in held-out loss relative to the empty graph}, with fewer than half of the adjacencies.

\begin{table}[t]
  \centering
  \small
  \caption{{Held-out pairwise prediction from the test sample. Adj. gives the number of adjacencies; $\widehat d_{\mathrm{eff}}$ is the estimated effective number of parameters defined in \eqref{eq:clbic_def}; Test NLL mean and SD summarize per-student negative log-likelihood (NLL) values; $\Delta\mathrm{NLL}$ is the mean NLL of each comparison model minus that of DAG-CLIP, with standard errors based on paired per-student differences in parentheses.}}
  \label{tab:timss_heldout}
  \resizebox{\linewidth}{!}{\begin{tabular}{lccccc}
\toprule
Model & Adj. & $\widehat d_{\mathrm{eff}}$ & Test NLL mean & SD & $10^3\,\Delta$NLL (SE) \\
\midrule
DAG-CLIP & 57 & 6.025 & 2.242957 & 0.3266 & reference \\
Saturated latent covariance & 120 & 6.124 & 2.242918 & 0.3266 & $-0.039$ ($0.020$) \\
Empty latent graph & 0 & 5.559 & 2.264818 & 0.3468 & $+21.861$ ($1.435$) \\
\bottomrule
\end{tabular}
}
\end{table}

The model-implied reliability of each latent construct ranges from 0.91 to 0.97, except for Home Educational Resources, where it is 0.28. This may be because three of its four indicators are binary and have nearly 100\% of responses in the category yes. Further measurement diagnostics are reported in the supplementary material.

\subsection{Background Knowledge and Interpretation}\label{subsec:timss_interpretation}

The single undirected edge {is found between ``Home Educational Resources'' (HER) and ``Digital Self-Efficacy'' (DSE), both from the general resources and experiences group with an estimated correlation of $0.70$. The restriction discussed in Section~\ref{subsec:timss_preparation} excludes arrows entering this group from the other groups and leaves within-group directions unrestricted.}  The remaining 56 directions are compelled {within the selected MEC}. 

Figure~\ref{fig:timss_summary}(B) {shows that the background variables HER, DSE, ``Sense of School Belonging'' (SSB), ``Bullying Exposure'' (BULL) and the context variable ``Confidence in Science'' (CS) point to science achievement (SA), and indirectly to mathematics achievement (MA) through SA.} The indirect effect of the background resources and experiences on mathematics achievement through science achievement suggests that their effects extend across subjects. The two achievement scores thus form one block linked to the background constructs. The direction SA $\to$ MA is compelled within the selected MEC.

Furthermore, achievement in mathematics and science may shape students' self-per\-ceptions and their perceived learning environment. In particular, MA $\to$ CM is consistent with stronger mathematics performance building confidence in mathematics, while SA $\to$ LLS suggests that achievement can encourage enjoyment of learning science. The edge CS $\to$ SA suggests that confidence may also contribute to science achievement. Thus, confidence plays different roles across the two subjects in the fitted model.

Finally, the paths from achievement to how students perceive disorderly behavior during Mathematics and Science Lessons (DML, DLS) are of interest. Their estimated standardized path coefficients obtained from the final joint refit of the selected CPDAG are given in brackets below next to each edge. MA $\to$ DML ($-0.654$), MA $\to$ DLS ($0.367$), SA $\to$ DLS ($-0.429$) and SA $\to$ DML ($0.457$) indicate that achievement is related to perceived disorderly behavior during mathematics and science lessons in different ways. The negative estimated coefficients for MA $\to$ DML and SA $\to$ DLS indicate that stronger achievement in mathematics or science is associated with less perceived classroom disorder in the corresponding subject. The positive associations found across subjects could be explained by high achievers in mathematics or science noticing more disruptive behavior in science or mathematics, respectively, due to boredom or less interest in the subject. 

\section{Discussion}\label{sec:dis}

In this work, we addressed the challenge of learning DAGs with latent constructs measured by non-Gaussian manifest variables, which are common in social and behavioral research. To avoid the heavy computational burden of traditional marginal likelihoods, we developed DAG-CLIP, a novel two-step composite-likelihood-based framework. The algorithm replaces the marginal likelihood with a pairwise likelihood, thereby avoiding high-dimensional integrals.
It first uses a smooth acyclicity-constrained optimization approach to screen out misspecified DAG structures, and then applies an information-criterion-guided backward-deletion procedure within the Markov equivalence class space to yield a sparse, refined causal graph. We established the theoretical consistency of DAG-CLIP in recovering the true Markov equivalence class, while extensive simulations and an application to large-scale educational survey data confirmed its practical effectiveness and scalability in handling complex measurement models.

The work can be extended in several directions. First, the current paper assumes unidimensional measurement models, i.e., each manifest variable is conditionally independent of the other latent variables given the single latent variable it measures. We make this assumption for ease of exposition, not as a technical requirement. In fact, as shown in \cite{KatsikatsouEtAl2012} and \cite{KatsikatsouMoustaki2016}, the pairwise likelihood function applies to complex multidimensional measurement designs, and the DAG-CLIP method can be extended to such settings.
{Second, the NOTEARS acyclicity constraint can be replaced by alternatives, such as the log-determinant constraint proposed in  \cite{bello2022dagma}. These alternatives preserve asymptotic consistency under the same regularity and numerical accuracy conditions, but may yield different finite-sample performance.}
{Third}, prior knowledge may sometimes be available about causal directions among a subset of variables. DAG-CLIP can easily incorporate such knowledge as constraints in both the screening and iterative pruning steps.
Finally, the method can be extended to exploratory settings where the relationship between manifest and latent variables is unknown and must be learned from data \citep[e.g.,][]{SilvaScheinesGlymourSpirtes2006,xie2020generalized,huang2022latent,DongEtAl2026LGES}.
In this case, one may combine recent representation learning methods \citep{RoheZeng2023,CuiChenOuyangXu2026}
and the proposed DAG-CLIP method to simultaneously define the latent variables via manifest variables and learn their causal relationships.

\clearpage
\renewcommand{\baselinestretch}{1.2}\selectfont
\setlength{\bibsep}{4pt} %
\bibliography{ref}

\begin{thebibliography}{74}
\providecommand{\natexlab}[1]{#1}
\providecommand{\url}[1]{\texttt{#1}}
\expandafter\ifx\csname urlstyle\endcsname\relax
  \providecommand{\doi}[1]{doi: #1}\else
  \providecommand{\doi}{doi: \begingroup \urlstyle{rm}\Url}\fi

\bibitem[Alfonzetti et~al.(2025{\natexlab{a}})Alfonzetti, Bellio, Chen, and
  Moustaki]{Alfonzetti2024}
Giuseppe Alfonzetti, Ruggero Bellio, Yunxiao Chen, and Irini Moustaki.
\newblock Pairwise stochastic approximation for confirmatory factor analysis of
  categorical data.
\newblock \emph{British Journal of Mathematical and Statistical Psychology},
  78\penalty0 (1):\penalty0 22--43, 2025{\natexlab{a}}.

\bibitem[Alfonzetti et~al.(2025{\natexlab{b}})Alfonzetti, Bellio, Chen, and
  Moustaki]{Alfonzetti2025composite}
Giuseppe Alfonzetti, Ruggero Bellio, Yunxiao Chen, and Irini Moustaki.
\newblock When composite likelihood meets stochastic approximation.
\newblock \emph{Journal of the American Statistical Association}, 120\penalty0
  (551):\penalty0 1906--1918, 2025{\natexlab{b}}.

\bibitem[Anderson and Rubin(1956)]{anderson1956statistical}
Theodore~W. Anderson and Herman Rubin.
\newblock Statistical inference in factor analysis.
\newblock In Jerzy Neyman, editor, \emph{Proceedings of the Third Berkeley
  Symposium on Mathematical Statistics and Probability}, volume~5, pages
  111--150. University of California Press, Berkeley, 1956.

\bibitem[Andersson et~al.(1997)Andersson, Madigan, and
  Perlman]{andersson1997characterization}
Steen~A. Andersson, David Madigan, and Michael~D. Perlman.
\newblock A characterization of {M}arkov equivalence classes for acyclic
  digraphs.
\newblock \emph{The Annals of Statistics}, 25\penalty0 (2):\penalty0 505--541,
  1997.

\bibitem[Aragam and Zhou(2015)]{aragam2015concave}
Bryon Aragam and Qing Zhou.
\newblock Concave penalized estimation of sparse {G}aussian {B}ayesian
  networks.
\newblock \emph{Journal of Machine Learning Research}, 16\penalty0
  (69):\penalty0 2273--2328, 2015.

\bibitem[Bartholomew et~al.(2011)Bartholomew, Knott, and
  Moustaki]{BartholomewKnottMoustaki2011}
David~J. Bartholomew, Martin Knott, and Irini Moustaki.
\newblock \emph{Latent Variable Models and Factor Analysis: A Unified
  Approach}.
\newblock Wiley, Chichester, 3rd edition, 2011.

\bibitem[Bello et~al.(2022)Bello, Aragam, and Ravikumar]{bello2022dagma}
Kevin Bello, Bryon Aragam, and Pradeep Ravikumar.
\newblock {DAGMA}: Learning {DAGs} via {M}-matrices and a log-determinant
  acyclicity characterization.
\newblock In \emph{Advances in Neural Information Processing Systems},
  volume~35, pages 8226--8239, 2022.

\bibitem[Bollen(1989)]{bollen1989structural}
Kenneth~A. Bollen.
\newblock \emph{Structural Equations with Latent Variables}.
\newblock Wiley Series in Probability and Mathematical Statistics. Wiley, New
  York, 1989.

\bibitem[Bollen and Pearl(2013)]{BollenPearl2013}
Kenneth~A. Bollen and Judea Pearl.
\newblock Eight myths about causality and structural equation models.
\newblock In Stephen~L. Morgan, editor, \emph{Handbook of Causal Analysis for
  Social Research}, pages 301--328. Springer, Dordrecht, 2013.

\bibitem[Cai et~al.(2019)Cai, Xie, Glymour, Hao, and Zhang]{cai2019triad}
Ruichu Cai, Feng Xie, Clark Glymour, Zhifeng Hao, and Kun Zhang.
\newblock Triad constraints for learning causal structure of latent variables.
\newblock In \emph{Advances in Neural Information Processing Systems},
  volume~32, pages 12883--12892, 2019.

\bibitem[Castelletti(2024)]{Castelletti2024}
Federico Castelletti.
\newblock Learning {Bayesian} networks: A copula approach for mixed-type data.
\newblock \emph{Psychometrika}, 89\penalty0 (2):\penalty0 658--686, 2024.

\bibitem[Chen et~al.(2024)Chen, Li, Shen, and Pan]{chen2024discovery}
Li~Chen, Chunlin Li, Xiaotong Shen, and Wei Pan.
\newblock Discovery and inference of a causal network with hidden confounding.
\newblock \emph{Journal of the American Statistical Association}, 119\penalty0
  (548):\penalty0 2572--2584, 2024.

\bibitem[Chen et~al.(2021)Chen, Dash, and Gao]{ChenDashGao2021}
Rui Chen, Sanjeeb Dash, and Tian Gao.
\newblock Integer programming for causal structure learning in the presence of
  latent variables.
\newblock In \emph{Proceedings of the 38th International Conference on Machine
  Learning}, volume 139 of \emph{Proceedings of Machine Learning Research},
  pages 1550--1560. PMLR, 2021.

\bibitem[Chen et~al.(2019)Chen, Drton, and Wang]{chen2019causal}
Wenyu Chen, Mathias Drton, and Y.~Samuel Wang.
\newblock On causal discovery with an equal-variance assumption.
\newblock \emph{Biometrika}, 106\penalty0 (4):\penalty0 973--980, 2019.

\bibitem[Chen et~al.(2025)Chen, Li, Liu, and Ying]{ChenLiLiuYing2025}
Yunxiao Chen, Xiaoou Li, Jingchen Liu, and Zhiliang Ying.
\newblock Item response theory---{A} statistical framework for educational and
  psychological measurement.
\newblock \emph{Statistical Science}, 40\penalty0 (2):\penalty0 167--194, 2025.

\bibitem[Chickering(2002{\natexlab{a}})]{chickering2002equivalence}
David~Maxwell Chickering.
\newblock Learning equivalence classes of {B}ayesian-network structures.
\newblock \emph{Journal of Machine Learning Research}, 2:\penalty0 445--498,
  2002{\natexlab{a}}.

\bibitem[Chickering(2002{\natexlab{b}})]{chickering2002optimal}
David~Maxwell Chickering.
\newblock Optimal structure identification with greedy search.
\newblock \emph{Journal of Machine Learning Research}, 3:\penalty0 507--554,
  2002{\natexlab{b}}.

\bibitem[Colombo and Maathuis(2014)]{colombo2014order}
Diego Colombo and Marloes~H. Maathuis.
\newblock Order-independent constraint-based causal structure learning.
\newblock \emph{Journal of Machine Learning Research}, 15\penalty0
  (116):\penalty0 3921--3962, 2014.

\bibitem[Colombo et~al.(2012)Colombo, Maathuis, Kalisch, and
  Richardson]{ColomboMaathuisKalischRichardson2012}
Diego Colombo, Marloes~H. Maathuis, Markus Kalisch, and Thomas~S. Richardson.
\newblock Learning high-dimensional directed acyclic graphs with latent and
  selection variables.
\newblock \emph{The Annals of Statistics}, 40\penalty0 (1):\penalty0 294--321,
  2012.

\bibitem[Cui et~al.(2026)Cui, Chen, Ouyang, and Xu]{CuiChenOuyangXu2026}
Chengyu Cui, Yunxiao Chen, Jing Ouyang, and Gongjun Xu.
\newblock Beyond vintage rotation: Bias-free sparse representation learning
  with oracle inference.
\newblock \emph{arXiv preprint arXiv:2602.22590}, 2026.

\bibitem[Dong et~al.(2025)Dong, Ng, Sun, Dai, Hao, Fan, Spirtes, Qiu, and
  Zhang]{DongEtAl2025MPRT}
Xinshuai Dong, Ignavier Ng, Boyang Sun, Haoyue Dai, Guang-Yuan Hao, Shunxing
  Fan, Peter Spirtes, Yumou Qiu, and Kun Zhang.
\newblock Permutation-based rank test in the presence of discretization and
  application in causal discovery with mixed data.
\newblock In \emph{Proceedings of the 42nd International Conference on Machine
  Learning}, volume 267 of \emph{Proceedings of Machine Learning Research},
  pages 14137--14151. PMLR, 2025.

\bibitem[Dong et~al.(2026)Dong, Ng, Dai, Sun, Song, Spirtes, and
  Zhang]{DongEtAl2026LGES}
Xinshuai Dong, Ignavier Ng, Haoyue Dai, Jiaqi Sun, Xiangchen Song, Peter
  Spirtes, and Kun Zhang.
\newblock Score-based greedy search for structure identification of partially
  observed causal models.
\newblock In \emph{International Conference on Learning Representations}, 2026.

\bibitem[Ellis and Wong(2008)]{ellis2008learning}
Byron Ellis and Wing~Hung Wong.
\newblock Learning causal {B}ayesian network structures from experimental data.
\newblock \emph{Journal of the American Statistical Association}, 103\penalty0
  (482):\penalty0 778--789, 2008.

\bibitem[Fishbein et~al.(2025)Fishbein, Taneva, and
  Kowolik]{FishbeinTanevaKowolik2025}
Bethany Fishbein, Milena Taneva, and Kamil Kowolik.
\newblock {TIMSS} 2023 {User Guide} for the {International Database}.
\newblock Technical report, {TIMSS \& PIRLS International Study Center, Boston
  College}, Chestnut Hill, MA, 2025.

\bibitem[Gao and Song(2010)]{GaoSong2010}
Xin Gao and Peter X.-K. Song.
\newblock Composite likelihood {Bayesian} information criteria for model
  selection in high-dimensional data.
\newblock \emph{Journal of the American Statistical Association}, 105\penalty0
  (492):\penalty0 1531--1540, 2010.

\bibitem[Gu and Xu(2019)]{GuXu2019}
Yuqi Gu and Gongjun Xu.
\newblock Learning attribute patterns in high-dimensional structured latent
  attribute models.
\newblock \emph{Journal of Machine Learning Research}, 20\penalty0
  (115):\penalty0 1--58, 2019.

\bibitem[Harris and Drton(2013)]{harris2013pc}
Naftali Harris and Mathias Drton.
\newblock {PC} algorithm for nonparanormal graphical models.
\newblock \emph{Journal of Machine Learning Research}, 14\penalty0
  (105):\penalty0 3365--3383, 2013.

\bibitem[Haughton(1988)]{haughton1988choice}
Dominique M.~A. Haughton.
\newblock On the choice of a model to fit data from an exponential family.
\newblock \emph{The Annals of Statistics}, 16\penalty0 (1):\penalty0 342--355,
  1988.

\bibitem[Hestenes(1969)]{Hestenes1969}
Magnus~R. Hestenes.
\newblock Multiplier and gradient methods.
\newblock \emph{Journal of Optimization Theory and Applications}, 4\penalty0
  (5):\penalty0 303--320, 1969.

\bibitem[Huang et~al.(2022)Huang, Low, Xie, Glymour, and
  Zhang]{huang2022latent}
Biwei Huang, Charles Jia~Han Low, Feng Xie, Clark Glymour, and Kun Zhang.
\newblock Latent hierarchical causal structure discovery with rank constraints.
\newblock In \emph{Advances in Neural Information Processing Systems},
  volume~35, pages 5549--5561, 2022.

\bibitem[{IEA}(2024)]{IEATIMSS2023G8}
{IEA}.
\newblock {TIMSS} 2023 {International Database}: {Trends in International
  Mathematics and Science Study}, {Grade} 8.
\newblock Data set, edition 1. International Association for the Evaluation of
  Educational Achievement and {TIMSS} \& {PIRLS} International Study Center,
  Boston College, 2024.

\bibitem[Kalisch and B{\"u}hlmann(2007)]{kalisch2007pc}
Markus Kalisch and Peter B{\"u}hlmann.
\newblock Estimating high-dimensional directed acyclic graphs with the
  {PC}-algorithm.
\newblock \emph{Journal of Machine Learning Research}, 8\penalty0
  (22):\penalty0 613--636, 2007.

\bibitem[Katsikatsou and Moustaki(2016)]{KatsikatsouMoustaki2016}
Myrsini Katsikatsou and Irini Moustaki.
\newblock Pairwise likelihood ratio tests and model selection criteria for
  structural equation models with ordinal variables.
\newblock \emph{Psychometrika}, 81\penalty0 (4):\penalty0 1046--1068, 2016.

\bibitem[Katsikatsou et~al.(2012)Katsikatsou, Moustaki, Yang-Wallentin, and
  J{\"o}reskog]{KatsikatsouEtAl2012}
Myrsini Katsikatsou, Irini Moustaki, Fan Yang-Wallentin, and Karl~G.
  J{\"o}reskog.
\newblock Pairwise likelihood estimation for factor analysis models with
  ordinal data.
\newblock \emph{Computational Statistics \& Data Analysis}, 56\penalty0
  (12):\penalty0 4243--4258, 2012.

\bibitem[Kummerfeld and Ramsey(2016)]{kummerfeld2016causal}
Erich Kummerfeld and Joseph Ramsey.
\newblock Causal clustering for 1-factor measurement models.
\newblock In \emph{Proceedings of the 22nd ACM SIGKDD International Conference
  on Knowledge Discovery and Data Mining}, pages 1655--1664. ACM, 2016.

\bibitem[Le~Thi and Pham~Dinh(2018)]{lethi2018dc}
Hoai~An Le~Thi and Tao Pham~Dinh.
\newblock {DC} programming and {DCA}: Thirty years of developments.
\newblock \emph{Mathematical Programming}, 169\penalty0 (1):\penalty0 5--68,
  2018.

\bibitem[Lee and Gu(2024)]{LeeGu2024}
Seunghyun Lee and Yuqi Gu.
\newblock Latent conjunctive {Bayesian} network: Unify attribute hierarchy and
  {Bayesian} network for cognitive diagnosis.
\newblock \emph{The Annals of Applied Statistics}, 18\penalty0 (3):\penalty0
  1988--2011, 2024.

\bibitem[Li et~al.(2023)Li, Shen, and Pan]{li2023inference}
Chunlin Li, Xiaotong Shen, and Wei Pan.
\newblock Inference for a large directed acyclic graph with unspecified
  interventions.
\newblock \emph{Journal of Machine Learning Research}, 24\penalty0
  (73):\penalty0 1--48, 2023.

\bibitem[Lindsay(1988)]{Lindsay1988}
Bruce~G. Lindsay.
\newblock Composite likelihood methods.
\newblock In Narahari~U. Prabhu, editor, \emph{Statistical Inference from
  Stochastic Processes}, volume~80 of \emph{Contemporary Mathematics}, pages
  221--239. American Mathematical Society, Providence, RI, 1988.

\bibitem[Liu et~al.(2022)Liu, Constantinou, and Guo]{liu2022measurement}
Yang Liu, Anthony~C. Constantinou, and Zhigao Guo.
\newblock Improving {B}ayesian network structure learning in the presence of
  measurement error.
\newblock \emph{Journal of Machine Learning Research}, 23\penalty0
  (324):\penalty0 1--28, 2022.

\bibitem[Luo et~al.(2021)Luo, Moffa, and Kuipers]{LuoMoffaKuipers2021}
Xiang~Ge Luo, Giusi Moffa, and Jack Kuipers.
\newblock Learning {Bayesian} networks from ordinal data.
\newblock \emph{Journal of Machine Learning Research}, 22\penalty0
  (266):\penalty0 1--44, 2021.

\bibitem[Meek(1995)]{meek1995causal}
Christopher Meek.
\newblock Causal inference and causal explanation with background knowledge.
\newblock In \emph{Proceedings of the Eleventh Conference on Uncertainty in
  Artificial Intelligence}, pages 403--410. Morgan Kaufmann, 1995.

\bibitem[Muth{\'e}n(1984)]{Muthen1984}
Bengt Muth{\'e}n.
\newblock A general structural equation model with dichotomous, ordered
  categorical, and continuous latent variable indicators.
\newblock \emph{Psychometrika}, 49\penalty0 (1):\penalty0 115--132, 1984.

\bibitem[Nandy et~al.(2018)Nandy, Hauser, and
  Maathuis]{nandy2018highdimensional}
Preetam Nandy, Alain Hauser, and Marloes~H. Maathuis.
\newblock High-dimensional consistency in score-based and hybrid structure
  learning.
\newblock \emph{The Annals of Statistics}, 46\penalty0 (6A):\penalty0
  3151--3183, 2018.

\bibitem[Ng et~al.(2020)Ng, Ghassami, and Zhang]{ng2020golem}
Ignavier Ng, AmirEmad Ghassami, and Kun Zhang.
\newblock On the role of sparsity and {DAG} constraints for learning linear
  {DAG}s.
\newblock In \emph{Advances in Neural Information Processing Systems},
  volume~33, pages 17943--17954, 2020.

\bibitem[Ng et~al.(2024)Ng, Dong, Dai, Huang, Spirtes, and
  Zhang]{NgEtAl2024SALAD}
Ignavier Ng, Xinshuai Dong, Haoyue Dai, Biwei Huang, Peter Spirtes, and Kun
  Zhang.
\newblock Score-based causal discovery of latent variable causal models.
\newblock In \emph{Proceedings of the 41st International Conference on Machine
  Learning}, volume 235 of \emph{Proceedings of Machine Learning Research},
  pages 37531--37552. PMLR, 2024.

\bibitem[Ni and Mallick(2022)]{ni2022ordinal}
Yang Ni and Bani Mallick.
\newblock Ordinal causal discovery.
\newblock In James Cussens and Kun Zhang, editors, \emph{Proceedings of the
  Thirty-Eighth Conference on Uncertainty in Artificial Intelligence}, volume
  180 of \emph{Proceedings of Machine Learning Research}, pages 1530--1540.
  PMLR, 2022.

\bibitem[Ni et~al.(2025)Ni, Chen, and Wang]{ni2025causal}
Yang Ni, Su~Chen, and Zeya Wang.
\newblock Causal structural modeling of survey questionnaires via a
  bootstrapped ordinal {B}ayesian network approach.
\newblock \emph{Psychometrika}, 90\penalty0 (1):\penalty0 229--250, 2025.

\bibitem[Nocedal and Wright(2006)]{nocedal2006numerical}
Jorge Nocedal and Stephen~J. Wright.
\newblock \emph{Numerical Optimization}.
\newblock Springer, New York, 2nd edition, 2006.

\bibitem[Olsson(1979)]{Olsson1979}
Ulf Olsson.
\newblock Maximum likelihood estimation of the polychoric correlation
  coefficient.
\newblock \emph{Psychometrika}, 44\penalty0 (4):\penalty0 443--460, 1979.

\bibitem[Pearl(2009)]{pearl2009causality}
Judea Pearl.
\newblock \emph{Causality: Models, Reasoning, and Inference}.
\newblock Cambridge University Press, Cambridge, 2nd edition, 2009.

\bibitem[Pearson(1900)]{Pearson1900}
Karl Pearson.
\newblock Mathematical contributions to the theory of evolution. {VII}. {On}
  the correlation of characters not quantitatively measurable.
\newblock \emph{Philosophical Transactions of the Royal Society of London.
  Series A}, 195:\penalty0 1--47, 1900.

\bibitem[Peng et~al.(2020)Peng, Shen, and Pan]{peng2020reconstruction}
Si~Peng, Xiaotong Shen, and Wei Pan.
\newblock Reconstruction of a directed acyclic graph with intervention.
\newblock \emph{Electronic Journal of Statistics}, 14\penalty0 (2):\penalty0
  4133--4164, 2020.

\bibitem[Peters and B{\"u}hlmann(2014)]{peters2014identifiability}
Jonas Peters and Peter B{\"u}hlmann.
\newblock Identifiability of {G}aussian structural equation models with equal
  error variances.
\newblock \emph{Biometrika}, 101\penalty0 (1):\penalty0 219--228, 2014.

\bibitem[Ramsey et~al.(2006)Ramsey, Zhang, and Spirtes]{ramsey2006adjacency}
Joseph Ramsey, Jiji Zhang, and Peter Spirtes.
\newblock Adjacency-faithfulness and conservative causal inference.
\newblock In \emph{Proceedings of the Twenty-Second Conference on Uncertainty
  in Artificial Intelligence}, pages 401--408. AUAI Press, 2006.

\bibitem[Rohe and Zeng(2023)]{RoheZeng2023}
Karl Rohe and Muzhe Zeng.
\newblock Vintage factor analysis with {V}arimax performs statistical
  inference.
\newblock \emph{Journal of the Royal Statistical Society Series B: Statistical
  Methodology}, 85\penalty0 (4):\penalty0 1037--1060, 2023.

\bibitem[Samejima(1969)]{Samejima1969}
Fumiko Samejima.
\newblock Estimation of latent ability using a response pattern of graded
  scores.
\newblock \emph{Psychometrika}, 34\penalty0 (S1):\penalty0 1--97, 1969.

\bibitem[Schwarz(1978)]{schwarz1978estimating}
Gideon Schwarz.
\newblock Estimating the dimension of a model.
\newblock \emph{The Annals of Statistics}, 6\penalty0 (2):\penalty0 461--464,
  1978.

\bibitem[Shen et~al.(2012)Shen, Pan, and Zhu]{shen2012likelihood}
Xiaotong Shen, Wei Pan, and Yunzhang Zhu.
\newblock Likelihood-based selection and sharp parameter estimation.
\newblock \emph{Journal of the American Statistical Association}, 107\penalty0
  (497):\penalty0 223--232, 2012.

\bibitem[Shimizu et~al.(2006)Shimizu, Hoyer, Hyv{\"a}rinen, and
  Kerminen]{shimizu2006linear}
Shohei Shimizu, Patrik~O. Hoyer, Aapo Hyv{\"a}rinen, and Antti Kerminen.
\newblock A linear non-{G}aussian acyclic model for causal discovery.
\newblock \emph{Journal of Machine Learning Research}, 7:\penalty0 2003--2030,
  2006.

\bibitem[Silva et~al.(2006)Silva, Scheines, Glymour, and
  Spirtes]{SilvaScheinesGlymourSpirtes2006}
Ricardo Silva, Richard Scheines, Clark Glymour, and Peter Spirtes.
\newblock Learning the structure of linear latent variable models.
\newblock \emph{Journal of Machine Learning Research}, 7:\penalty0 191--246,
  2006.

\bibitem[Spirtes et~al.(2000)Spirtes, Glymour, and
  Scheines]{SpirtesGlymourScheines2000}
Peter Spirtes, Clark~N. Glymour, and Richard Scheines.
\newblock \emph{Causation, Prediction, and Search}.
\newblock The MIT Press, Cambridge, MA, 2nd edition, 2000.

\bibitem[Taeb et~al.(2024)Taeb, Gamella, Heinze-Deml, and
  B{\"u}hlmann]{taeb2024latent}
Armeen Taeb, Juan~L. Gamella, Christina Heinze-Deml, and Peter B{\"u}hlmann.
\newblock Learning and scoring {G}aussian latent variable causal models with
  unknown additive interventions.
\newblock \emph{Journal of Machine Learning Research}, 25\penalty0
  (293):\penalty0 1--68, 2024.

\bibitem[van~de Geer and B{\"u}hlmann(2013)]{vandegeer2013l0}
Sara van~de Geer and Peter B{\"u}hlmann.
\newblock $\ell_0$-penalized maximum likelihood for sparse directed acyclic
  graphs.
\newblock \emph{The Annals of Statistics}, 41\penalty0 (2):\penalty0 536--567,
  2013.

\bibitem[Varin et~al.(2011)Varin, Reid, and Firth]{VarinReidFirth2011}
Cristiano Varin, Nancy Reid, and David Firth.
\newblock An overview of composite likelihood methods.
\newblock \emph{Statistica Sinica}, 21\penalty0 (1):\penalty0 5--42, 2011.

\bibitem[von Davier et~al.(2024)von Davier, Fishbein, and
  Kennedy]{vonDavierFishbeinKennedy2024}
Matthias von Davier, Bethany Fishbein, and Ann~M. Kennedy, editors.
\newblock \emph{{TIMSS} 2023 Technical Report: Methods and Procedures}.
\newblock {TIMSS \& PIRLS International Study Center, Boston College}, Chestnut
  Hill, MA, 2024.

\bibitem[Wang et~al.(2024)Wang, Shen, and Pan]{wang2024peeling}
Minjie Wang, Xiaotong Shen, and Wei Pan.
\newblock Causal discovery with generalized linear models through peeling
  algorithms.
\newblock \emph{Journal of Machine Learning Research}, 25\penalty0
  (310):\penalty0 1--49, 2024.

\bibitem[Wu and Estabrook(2016)]{WuEstabrook2016}
Hao Wu and Ryne Estabrook.
\newblock Identification of confirmatory factor analysis models of different
  levels of invariance for ordered categorical outcomes.
\newblock \emph{Psychometrika}, 81\penalty0 (4):\penalty0 1014--1045, 2016.

\bibitem[Xie et~al.(2020)Xie, Cai, Huang, Glymour, Hao, and
  Zhang]{xie2020generalized}
Feng Xie, Ruichu Cai, Biwei Huang, Clark Glymour, Zhifeng Hao, and Kun Zhang.
\newblock Generalized independent noise condition for estimating latent
  variable causal graphs.
\newblock In \emph{Advances in Neural Information Processing Systems},
  volume~33, pages 14891--14902, 2020.

\bibitem[Yu et~al.(2019)Yu, Chen, Gao, and Yu]{yu2019dag}
Yue Yu, Jie Chen, Tian Gao, and Mo~Yu.
\newblock {DAG-GNN}: {DAG} structure learning with graph neural networks.
\newblock In \emph{Proceedings of the 36th International Conference on Machine
  Learning}, volume~97 of \emph{Proceedings of Machine Learning Research},
  pages 7154--7163. PMLR, 2019.

\bibitem[Yuan et~al.(2019)Yuan, Shen, Pan, and Wang]{yuan2019constrained}
Yiping Yuan, Xiaotong Shen, Wei Pan, and Zizhuo Wang.
\newblock Constrained likelihood for reconstructing a directed acyclic
  {G}aussian graph.
\newblock \emph{Biometrika}, 106\penalty0 (1):\penalty0 109--125, 2019.

\bibitem[Zhang et~al.(2026)Zhang, Wang, and Gu]{zhang2026discrete}
Wenjin Zhang, Yixin Wang, and Yuqi Gu.
\newblock Discrete causal representation learning.
\newblock \emph{arXiv preprint arXiv:2603.25017}, 2026.

\bibitem[Zhao et~al.(2022)Zhao, He, and Wang]{zhao2022nonGaussian}
Ruixuan Zhao, Xin He, and Junhui Wang.
\newblock Learning linear non-{G}aussian directed acyclic graph with diverging
  number of nodes.
\newblock \emph{Journal of Machine Learning Research}, 23\penalty0
  (269):\penalty0 1--34, 2022.

\bibitem[Zheng et~al.(2018)Zheng, Aragam, Ravikumar, and Xing]{ZhengEtAl2018}
Xun Zheng, Bryon Aragam, Pradeep~K. Ravikumar, and Eric~P. Xing.
\newblock {DAGs with NO TEARS}: Continuous optimization for structure learning.
\newblock In \emph{Advances in Neural Information Processing Systems},
  volume~31, pages 9472--9483, 2018.

\end{thebibliography}
\end{document}